\documentclass[a4paper,12pt]{article}
\usepackage{xcolor}
\usepackage{jheppub}
\usepackage{epstopdf}
\usepackage{graphicx}
\usepackage{epsfig}
\usepackage{dcolumn}  
\usepackage{bm}    
\usepackage{amssymb} 
\usepackage{amsmath,bm}
\usepackage{amsfonts}    
\usepackage{slashed}  
\usepackage{youngtab}
\usepackage{pgf,tikz}
\usepackage{tkz-euclide}
\usetkzobj{all}
\usepackage{pgfplots}

\pgfplotsset{compat=1.8}
\usetikzlibrary{arrows, decorations.pathmorphing, calc,intersections,through,backgrounds,patterns,fit,arrows.meta}

\usetikzlibrary{arrows}
\usepackage[mathscr]{euscript}
\usepackage{epsfig}
\usepackage{pdfpages}
\usepackage{verbatim}
\usepackage{tensor}

\allowdisplaybreaks[1]

\title{\boldmath $\mathcal{N}=(3,3)$ holography on AdS$_3\times$(S$^3\times$S$^3\times$S$^1$)/${\mathbb Z_2}$}

\author[a]{Lorenz Eberhardt}
\author[b]{and Ida G. Zadeh}

\affiliation[a]{Institut f\"ur Theoretische Physik, ETH Z\"urich,\\CH-8093 Z\"urich, Switzerland}
\affiliation[b]{Department of Mathematics, ETH Z\"urich,\\CH-8092 Z\"urich, Switzerland}

\emailAdd{eberhardtl@itp.phys.ethz.ch}
\emailAdd{zadeh@math.ethz.ch}

\newcommand{\be}{\begin{equation}}
\newcommand{\ee}{\end{equation}}
\newcommand{\bea}{\begin{eqnarray}}
\newcommand{\eea}{\end{eqnarray}}
\newcommand{\bb}{\mathbb}

\def\2{{$\mathcal N=2$ }}
\def\3{{$\mathcal N=3$ }}

\def\l{{large $\mathcal N=4$ }}
\def\sk{{$\mathcal S_\kappa$ }}
\def\symsk{{$\mathrm{Sym}^N(\mathcal S_\kappa)$}}
\def\s0{{$\mathcal S_0$ }}
\def\syms0{{$\mathrm{Sym}^N(\mathcal S_0)$}}

\abstract{We consider string theory on AdS$_3$ $\times$ (S$^3$ $\times$ S$^3$ $\times$ S$^1)/\mathbb Z_2$, a background supporting $\mathcal N=(3,3)$ spacetime supersymmetry. We propose that string theory on this background is dual to the symmetric product orbifold of $\mathcal S_0/\mathbb Z_2$ where $\mathcal S_0$ is a theory of four free fermions and one free boson. We show that the BPS spectra of the two sides of the duality match precisely. Furthermore, we compute the elliptic genus of the dual CFT and that of the supergravity limit of string theory and demonstrate that they match, hence providing non-trivial support for the holographic proposal.}

\begin{document} 
\maketitle
\flushbottom

\section{Introduction}\label{intro}
New instances of the stringy holographic correspondence on AdS$_3$ backgrounds have been recently formulated. On the one hand, a large $\mathcal N=(4,4)$ duality is proposed in \cite{Eberhardt:2017pty} for type IIB string theory on  ${\rm AdS}_3\times{\rm S}^3\times{\rm S}^3\times{\rm S}^1$, following up on earlier developments \cite{Eberhardt:2017fsi,Baggio:2017kza,Gukov:2004ym,deBoer:1999gea,Elitzur:1998mm,Boonstra:1998yu}. The dual CFT is conjectured to be the symmetric product orbifold of the so-called $\mathcal S_\kappa$ CFTs which describe the $\sigma$-model on ${\rm S}^3\times {\rm S}^1$. On the other hand, a novel family of $\mathcal N=(2,2)$ dualities is proposed in \cite{Datta:2017ert, Eberhardt:2017uup} through considering type IIB string theory on quotient backgrounds ${\rm AdS}_3\times({\rm S}^3\times\mathbb T^4)/{\rm D}_n$ and ${\rm AdS}_3 \times ({\rm S}^3 \times {\rm K3})/\mathbb{Z}_2$, where ${\rm D}_n$ is the dihedral group. In these constructions, the original small $\mathcal N=(4,4)$ supersymmetry supported by $\mathbb T^4$ and K3 \cite{Maldacena:1997re} is reduced to $\mathcal N=(2,2)$ supersymmetry. The conjectured dual CFTs are again symmetric product orbifolds and the seed theories are CFTs on $\mathbb T^4/{\rm D}_n$ and $\mathrm{K3}/\mathbb{Z}_2$, respectively.

Motivated by these developments, in this work we study another example of the AdS$_3$/CFT$_2$ duality with non-maximal supersymmetry. We consider the construction of Yamaguchi et al.~\cite{Yamaguchi:1999gb} and study string theory on the orbifold background ${\rm AdS}_3\times({\rm S}^3\times{\rm S}^3\times{\rm S}^1)/\mathbb{Z}_2$. The action of the $\mathbb Z_2$ orbifold is implemented by exchanging the two three-spheres and reflecting the circle ${\rm S}^1$. This action may be realised such that the spacetime supersymmetry is reduced from large $\mathcal N=4$ to either $\mathcal N=3$ or $\mathcal N=1$. We consider the former and study string theory configuration with $\mathcal N=(3,3)$ supersymmetry. We conjecture that the CFT dual to this configuration is the symmetric orbifold of $\mathcal S_0/\mathbb Z_2$, where $\mathcal S_0$ (a member of the $\mathcal S_\kappa$ family with $\kappa=0$) is a theory of four free fermions and one free boson.

We compute the BPS spectra of the string worldsheet theory and of the dual CFT and find that they match precisely. We provide further non-trivial support for the proposed duality by matching a supersymmetric index between the bulk and boundary theories. The elliptic genus of the $\mathcal N=3$ CFTs vanishes due to the presence of a fermionic zero mode. We define a non-vanishing index in the NS sector, which is composed of chiral primaries in the right-moving sector and arbitrary excitations in the left-moving sector, compute it on both sides of the duality, and find that they match. This provides additional non-trivial evidence for the proposed holographic duality.

The plan of this paper is as follows. In section \ref{duality} we discuss the realisation of the action of the $\mathbb Z_2$ orbifold and propose the associated AdS$_3$/CFT$_2$ duality with $\mathcal N=(3,3)$ supersymmetry. In section \ref{CFT_bps} we compute the BPS spectrum of the proposed dual CFT. We define and compute a modified elliptic genus for the dual CFT in section \ref{CFT_ellgen}. We next analyse the string worldsheet theory in terms of the WZW models in section \ref{sheet_bps} and derive its BPS spectrum. We find that the spectrum precisely matches that of the dual CFT. In section \ref{sheet_ellgen} we compute the modified elliptic genus of the worldsheet theory in the supergravity limit and show that it reproduces the corresponding CFT results. Finally, in section \ref{conc} we conclude and discuss future directions of research. We present the details of the $\mathcal S_0$ CFT in appendix \ref{app_S0}. The large $\mathcal N=4$ and $\mathcal N=3$ algebras and the associated characters are reviewed in appendix \ref{app_alg}. Some technical details of the symmetric orbifold computations are presented in appendix \ref{app_SymN}. Finally, Jacobi theta function identities that we use are outlined  in appendix \ref{app_theta}.

\section{String theory on ${\rm AdS}_3 \times {\rm S}^3 \times {\rm S}^3 \times {\rm S}^1$}\label{duality}
The holographic duality for type IIB string theory on the ${\rm AdS}_3 \times {\rm S}^3 \times {\rm S}^3 \times {\rm S}^1$ background \cite{Eberhardt:2017pty} proposes that the dual CFT is the symmetric product orbifold of \sk theories, $\mathrm{Sym}^N(\mathcal S_\kappa)$. The \sk CFTs are supersymmetric $\sigma$-models on ${\rm S}^3 \times {\rm S}^1$ and have large $\mathcal N=(4,4)$ superconformal symmetry \cite{Eberhardt:2017pty,Gukov:2004fh}. The R-symmetry is $\mathfrak{su}(2)^+\oplus\mathfrak{su}(2)^-\oplus \mathfrak{u}(1)$ for both the left- and right-moving sectors. The BPS spectra of \symsk, the worldsheet theory described in terms of WZW models associated to ${\rm AdS}_3 \times {\rm S}^3 \times {\rm S}^3 \times {\rm S}^1$, and supergravity on this background were computed in \cite{Eberhardt:2017pty,Eberhardt:2017fsi} and it was shown that the spectra of the three theories match precisely.

In this work we consider the construction of \cite{Yamaguchi:1999gb} and study string theory on the $\mathrm{AdS}_3\times(\mathrm S^3\times\mathrm S^3\times\mathrm S^1)/\mathbb Z_2$ orbifold background. The action of the $\mathbb Z_2$ orbifold is realised by exchanging the two three-spheres and simultaneously reflecting the circle. This amounts to exchanging the two affine $\mathfrak{su}(2)^\pm$ algebras. The $\mathbb Z_2$ action then imposes the condition that the two spheres have equal radii, i.e. that the levels of the two affine $\mathfrak{su}(2)$'s coincide: $k^+=k^-$. This corresponds to $\kappa=0$.

The diagonal $\mathfrak{su}(2)$ of the R-symmetry algebra survives the orbifold projection. The currents of the diagonal $\mathfrak{su}(2)$ are the sum of the currents of the two $\mathfrak{su}(2)^\pm$ affine algebras and generate the R-symmetry of the worldsheet theory of string theory on the $\mathrm{AdS}_3\times(\mathrm S^3\times\mathrm S^3\times\mathrm S^1)/\mathbb Z_2$ background. Before taking the orbifold projection, the four left-moving supercurrents of the original large $\mathcal N=4$ SCA transform in the representation $\mathbf{(2,2)}$ of the $\mathfrak{su}(2)^+\oplus\mathfrak{su}(2)^-$. After taking the $\mathbb Z_2$-orbifold, the supercurrents transform in the $\mathbf{3}\oplus \mathbf{1}$ representations of the diagonal $\mathfrak{su}(2)$ (and likewise, the same result holds for the right-moving supercurrents). The action of the $\mathbb Z_2$ can be taken such that either the triplet or the singlet supercurrents survive the projection. The two operations are then shown in \cite{Yamaguchi:1999gb} to reduce the spacetime supersymmetry and yield $\mathcal N=3$ and $\mathcal N=1$ superconformal symmetry, respectively (see section \ref{sheet_bps} for more details). Taking into account the contribution from the right-moving part, the theory admits $\mathcal N=(3,3)$, $\mathcal N=(3,1)$, $\mathcal N=(1,3)$, or $\mathcal N=(1,1)$ supersymmetry.

Working within the context of the large $\mathcal N=(4,4)$ AdS$_3$/CFT$_2$ correspondence for the $\mathrm{AdS}_3\times\mathrm S^3\times\mathrm S^3\times\mathrm S^1$ background \cite{Eberhardt:2017pty} and performing the $\mathbb Z_2$ orbifold action of \cite{Yamaguchi:1999gb} described above, we propose that string theory on $\mathrm{AdS}_3\times(\mathrm S^3\times\mathrm S^3\times\mathrm S^1)/\mathbb Z_2$ background with $\mathcal N=(3,3)$ supersymmetry is dual to the symmetric product orbifold $\mathrm{Sym}^N(\mathcal S_0/\mathbb Z_2)$.

\subsection{Fluxes and charges}\label{duality_flux}
Let us briefly describe the fluxes of the model. For $\mathrm{AdS}_3 \times \mathrm{S}^3_+ \times \mathrm{S}^3_-\times \mathrm{S}^1$, there are three charges, which we denote by $Q_1^0$, $Q_5^{+,0}$ and $Q_5^{-,0}$. These correspond to the number of D1, $\mathrm{D5}^+$ and $\mathrm{D5}^-$-branes in the brane construction (or F1-strings and NS5-branes in the NS-NS background we are treating). $\mathrm{S}^3_\pm$ is supported by $Q_5^{\pm,0}$ units of flux, whereas $Q_1^0$ is related to the radius of the ${\rm S}^1$, see \cite[eq. (2.44)]{Gukov:2004ym}.

When performing the ${\bb Z}_2$-orbifold, there is an associated map in cohomology, which maps the fluxes of the quotient geometry to the original geometry. In this case, the two three-spheres get interchanged, which results in the quotient space to have only one value of $\mathrm{D5}$-brane (or NS5-brane) charge.\footnote{This is reflected in $\mathrm{H}^3((\mathrm{S}^3 \times \mathrm{S}^3 \times \mathrm{S}^1)/\mathbb{Z}_2;\mathbb{R}) \cong \mathbb{R}$.} The fluxes of the quotient geometry are then related to the original fluxes as
\be 
2Q_1=Q_1^0\ , \quad Q_5=Q_5^{+,0}=Q_5^{-,0}\ .
\ee
Note that we can perform the quotient only when $Q_5^{+,0}=Q_5^{-,0}$, i.e.~when the three-spheres have equal size. The factor of 2 in the D1-brane charge comes from the fact that the quotient map is a degree 2 map.

Let us also comment on the Brown-Henneaux central charge \cite{Brown:1986nw}. For the unorbifolded geometry, it is given by \cite{deBoer:1999gea, Gukov:2004fh, Tong:2014yna}:
\be 
c^0=\frac{6Q_1^0 Q_5^{+,0}Q_5^{-,0}}{Q_5^{+,0}+Q_5^{-,0}}=3 Q_1^0 Q_5^{+,0}\ .
\ee
The classical Brown-Henneaux central charge formula for the background $\mathrm{AdS}_3 \times \mathcal{M}_7$ is given by
\be 
c=\frac{3 \ell}{2G_{3\mathrm{d}}}\ ,\quad G_{3\mathrm{d}}=\frac{G_{10\mathrm{d}}}{\mathrm{vol}(\mathcal{M}_7)}\ .
\ee
where $\ell$ is the $\mathrm{AdS}_3$-radius and $G_{3\mathrm{d}}$ and $G_{10\mathrm{d}}$ are the 3-dimensional and 10-dimensional Newton constants, respectively. $\ell$ and $G_{10\mathrm{d}}$ are invariant under the orbifold action but the volume of $\mathcal{M}_7$ is only half as large after the orbifolding procedure, so we conclude that the Brown-Henneaux central charge of the orbifold geometry is given by
\be 
c=\frac{c^0}{2}=\frac{3}{2}Q_1^0 Q_5^{+,0}=3 Q_1 Q_5\ . \label{Brown Henneaux central charge}
\ee
There might be order one corrections to this formula, as for the Brown-Henneaux central charge on ${\rm AdS}_3 \times {\rm S}^3 \times {\rm K3}$ \cite{Aharony:1999ti, Beccaria:2014qea}.

The theory ${\cal S}_0/{\bb Z}_2$ has central charge 3, hence we conjecture that the dual CFT to the background ${\rm AdS}_3 \times ({\rm S}^3 \times {\rm S}^3 \times {\rm S}^1)/\mathbb{Z}_2$ is
\be 
\mathrm{Sym}^{Q_1Q_5}({\cal S}_0/\mathbb{Z}_2)\ . \label{number of copies}
\ee

\section{BPS spectrum of the dual CFT}\label{CFT_bps}
In this section we first review the BPS spectrum of the $\mathcal S_0$ theory in subsection \ref{CFT_bps_S0}. We next discuss in subsection \ref{CFT_bps_Z2} the actions of the $\mathbb Z_2$ orbifold on the $\mathcal S_0$ CFT which yield $\mathcal N=3$ and $\mathcal N=1$ superconformal symmetries. We note that only the reduction to the $\mathcal N=3$ SCA has an $\mathcal N=2$ subalgebra, and hence, a BPS spectrum. We derive this BPS spectrum in section \ref{CFT_bps_S0Z2}. For the $\mathcal N=1$ theory, we derive the spectrum of the theory which survives the action of the orbifold. In section \ref{CFT_bps_symS0Z2} we compute  the BPS spectrum of the symmetric product orbifold $\mathrm{Sym}^N(\mathcal S_0/\mathbb Z_2)$ with $\mathcal N=(3,3)$ supersymmetry. Finally, in section \ref{CFT_bps_moduli} we discuss the moduli of the dual CFT. We present the details of the analyses of the symmetric orbifold in appendix \ref{app_SymN_bps}.

\subsection{BPS spectrum of $\mathcal S_0$}\label{CFT_bps_S0}
Let us recall the definition of the so-called \s0 theory \cite{Sevrin:1988ew, Gukov:2004ym, Eberhardt:2017pty}. It consists of four free fermions $\psi^{\mu\nu}$ with $\mu, \nu \in \{+,-\}$ and one free boson $\partial \phi$.  These generating fields give the R-symmetry $\mathfrak{su}(2)_1 \oplus \mathfrak{su}(2)_1 \oplus \mathfrak{u}(1)$ of the large ${\cal N}=4$ SCA, as reviewed in more detail in Appendix~\ref{app_S0} (the subscripts denotes the level of the algebra). We denote these two affine $\mathfrak{su}(2)_1$-algebras in the following as $\mathfrak{su}(2)^+_1$ and $\mathfrak{su}(2)^-_1$. $\mu$ and $\nu$ are then bispinor indices of the $\mathfrak{su}(2)_1 \oplus \mathfrak{su}(2)_1$-algebra.

The BPS spectrum of the \s0 theory is derived e.g.~in \cite{Eberhardt:2017pty}. The only BPS states in the charge-zero sector are the vacuum together with their superconformal descendants:\footnote{Here and in the following, a tilde denotes the right-movers.}
\be 
|0 \rangle\ , \quad \psi_{-\frac{1}{2}}^{++} |0 \rangle\ , \quad \tilde{\psi}_{-\frac{1}{2}}^{++} |0 \rangle\ , \quad \psi^{++}_{-\frac{1}{2}}\tilde{\psi}_{-\frac{1}{2}}^{++} |0 \rangle
\ee
Hence the corresponding Hodge-diamond is of the form
\be\label{hdgT2}
\begin{tabular}{ l c r }
  {} & 1 & {} \\
  1 & {} & 1 \ ,\\
  {} & 1 & {} \\
\end{tabular}
\ee
which is the same as the Hodge diamond of the sigma model on $\mathbb T^2$, see \cite[section 5.3]{Eberhardt:2017pty}.

\subsection{Action of the $\mathbb Z_2$ orbifold}\label{CFT_bps_Z2}
The generators of the \l algebra can be realised in terms of the free fields{\footnote{We thank Matthias Gaberdiel for sharing a note with us on the explicit form of this realisation.} as presented in appendix \ref{app_S0} (see also \cite{Schoutens:1987uu,Gukov:2004ym}).

We would like to determine the action of the $\mathbb Z_2$-orbifold on the generators of the large $\mathcal N=4$ algebra such that it is reduced to the $\mathcal N=3$ or $\mathcal N=1$ SCA. As discussed in section \ref{duality}, the $\mathbb Z_2$-action exchanges the two affine $\mathfrak{su}(2)_1$ algebras of the R-symmetry algebra and thus imposes the condition $k^+=k^-=1$ (hence, $\kappa=0$) on the levels. $\mathbb Z_2$ also acts by inverting the sign of the free boson:
\be\label{Z2boson}
\partial \phi\longmapsto-\partial \phi\ .
\ee
For the four free fermions, we consider the following two actions:
\bea
&&\!\!\!\!\!\!\!\!(i)\quad\psi^{++}\longmapsto-\psi^{++},\quad\psi^{--} \longmapsto-\psi^{--},\quad\psi^{+-}\longmapsto\psi^{-+},\quad\psi^{-+}\longmapsto\psi^{+-},\label{casei}\quad\\
&&\!\!\!\!\!\!\!\!(ii)\quad\!\!\psi^{++}\longmapsto\psi^{++},\quad\quad\!\!\psi^{--} \longmapsto\psi^{--},\quad\;\;\;\psi^{+-}\longmapsto-\psi^{-+},\quad\!\!\!\!\psi^{-+}\longmapsto-\psi^{+-}.\label{caseii}\quad
\eea
Using eq.s (\ref{S0_A10})-(\ref{S0_GTUQ16}), one can check that in both cases the two $\mathfrak{su}(2)_1^\pm$ algebras are indeed exchanged and that the stress-energy tensor is preserved. Let us consider the transformation of the supercurrents. For case ($i$) we find:
\be\label{Gscasei_i}
G^{++}\longmapsto G^{++},\qquad G^{--}\longmapsto G^{--},\qquad G^{+-}\longmapsto G^{-+},\qquad G^{-+}\longmapsto G^{+-}.
\ee
The supercharges $G^{++}$, $G^{--}$, and the combination $G^{+-}+G^{-+}$ are preserved under the $\mathbb Z_2$-action. The large $\mathcal N=4$ SCA is reduced to the $\mathcal N=3$ algebra in this case \cite{Miki:1989ri}. The generators of the algebra and their (anti-)commutation relations as derived from this reduction are presented in appendix \ref{app_alg_N3}. 

For case ($ii$) in eq. (\ref{caseii}) we have:
\be\label{Gscaseii_i}
G^{++}\longmapsto-G^{++},\quad G^{--}\longmapsto-G^{--},\quad G^{+-}\longmapsto-G^{-+},\quad G^{-+}\longmapsto-G^{+-},
\ee
where only the combination $G^{+-}-G^{-+}$ survives the $\mathbb Z_2$ action. The large $\mathcal N=4$ SCA is reduced to the $\mathcal N=1$ SCA in this case.\footnote{In fact, to a semi-direct product of the $\mathcal N=1$ SCA with the diagonal affine $\mathfrak{su}(2)_2$-algebra, but this will not be important in what follows.} One can check that (\ref{casei}) and (\ref{caseii}) are the only two consistent $\mathbb Z_2$-actions which exchange the two $\mathfrak{su}(2)^\pm_1$ algebras.

\subsection{BPS spectrum of $\mathcal S_0/\mathbb Z_2$}\label{CFT_bps_S0Z2}
In this section, we derive the BPS spectrum of the $\mathcal{S}_0/\mathbb{Z}_2$ CFT. 

Let us start by computing the NS sector partition function for the $\mathcal N=3$ theory in (\ref{casei}). We recall that, under the $\mathbb Z_2$ action, the R-symmetry of the theory is reduced to the diagonal $\mathfrak{su}(2)_2$ algebra. We introduce a chemical potential, $z$, to keep track of the $\mathfrak{u}(1)$-charge (resp.~the $\mathfrak{su}(2)$-spin). Two of the four free fermions, $\psi^{++}$ and $\psi^{--}$, are charged under the $\mathfrak{u}(1)$ whereas the other two, $\psi^{+-}$ and $\psi^{-+}$, are not. Moreover, according to eq.~(\ref{casei}), the two charged fermions, $\psi^{++}$ and $\psi^{--}$, and the linear combination $\psi^{+-}+\psi^{-+}$ of uncharged fermions are flipped under the $\mathbb Z_2$ and are orbifold odd. All in all, the contribution to the partition function before orbifolding is of the form
\be\label{unt_unp}
\frac{1}{2}\Theta(\tau;\bar\tau;R)\left|\frac{\vartheta_3(z|\tau)\vartheta_3(\tau)}{\eta(\tau)^3}\right|^2\ ,
\ee
where $\Theta(\tau;\bar\tau;R)$ represents the theta-function associated to the compact free boson. We already included the factor of $\tfrac{1}{2}$, which is the usual factor $1/|\mathbb{Z}_2|$ introduced by the projection operator. A pair of charged fermions contribute a factor of $\big|{\vartheta_3(z|\tau)}/{\eta(\tau)}\big|^2$ and each uncharged fermion contributes a factor of $\big|{\vartheta_3(\tau)^{\frac12}}/{\eta(\tau)^{\frac12}}\big|^2$.

We next consider the contribution from the orbifold projected states in the untwisted sector as well as the unprojected and projected states in the twisted sector of the $\mathbb Z_2$. The details of the derivation may be found in  \cite[section 10.4.3]{DiFrancesco:1997nk}, the formula follows essentially by requiring invariance under modular transformations. The full NS sector partition function reads:
\begin{align}\label{Z_NS_S0Z2}
Z_{\mathrm{NS}}(z,\tau;\bar z,\bar\tau)&=\frac{1}{2}\Theta(\tau;\bar\tau;R)\left|\frac{\vartheta_3(z|\tau)\vartheta_3(\tau)}{\eta(\tau)^3}\right|^2+\left|\frac{\vartheta_4(z|\tau)\vartheta_3(\tau)^{\frac{1}{2}}\vartheta_4(\tau)^{\frac{1}{2}}}{\vartheta_2(\tau)^{\frac{1}{2}}\eta(\tau)^{\frac{3}{2}}}\right|^2\nonumber\\
&\qquad+\left|\frac{\vartheta_2(z|\tau)\vartheta_3(\tau)^{\frac{1}{2}}\vartheta_2(\tau)^{\frac{1}{2}}}{\vartheta_4(\tau)^{\frac{1}{2}}\eta(\tau)^{\frac{3}{2}}}\right|^2+\left|\frac{\vartheta_1(z|\tau)\vartheta_3(\tau)^{\frac{1}{2}}\vartheta_1(\tau)^{\frac{1}{2}}}{\vartheta_3(\tau)^{\frac{1}{2}}\eta(\tau)^{\frac{3}{2}}}\right|^2\ .
\end{align}
The last term formally vanishes, but it is useful to keep it as a book-keeping device. We note that the $\vartheta_3(\tau)^{\frac12}$ factor in all four terms of the above equation corresponds to the uncharged fermion which is not affected by the $\mathbb Z_2$ orbifold: it is just a bystander.

We shall now read off the BPS spectrum of the theory in the NS sector. Only the zero modes of $\vartheta_1$ and $\vartheta_2$ and the one-half modes of $\vartheta_3$ and $\vartheta_4$ have a chance to produce BPS states. Including only those, the partition function reduces to:
\begin{align}\label{S0Z2_bps}
|q|^{\frac{1}{4}} Z_{\mathrm{NS}}(z,\tau;\bar z,\bar\tau)&=\tfrac{1}{2}(1+y q^{\frac{1}{2}})(1+\bar{y}\bar{q}^{\frac{1}{2}})+\tfrac{1}{2}(1-y q^{\frac{1}{2}})(1-\bar{y}\bar{q}^{\frac{1}{2}})\nonumber\\
&\quad+2q^{\frac{1}{4}}\bar{q}^{\frac{1}{4}}(y^{\frac{1}{2}}+y^{-\frac{1}{2}})(y^{\frac{1}{2}}+y^{-\frac{1}{2}})+\text{non BPS}\nonumber \\
&=1+2q^{\frac{1}{4}}\bar{q}^{\frac{1}{4}}y^{\frac{1}{2}}\bar{y}^{\frac{1}{2}}+ q^{\frac{1}{2}}\bar{q}^{\frac{1}{2}}y\bar{y}+\text{non BPS}\ .
\end{align}
Thus, the corresponding Hodge-diamond is
\be\label{S0Z2hd}
\begin{tabular}{ccc}
& 1 & \\
0 & 2 & 0 \\
& 1 &
\end{tabular}\ .
\ee
We extend the notion of Hodge-diamond to incorporate also half-integer cohomology, see \cite{Datta:2017ert}. The corresponding Hodge-numbers will be denoted by $h_{p,q}$ with $p,q  \in \tfrac{1}{2}\mathbb{Z}$. The requirement of half-integer spin implies that $h_{p,q}=0$, unless $p+q \in \mathbb{Z}$.

We could have also derived the above BPS spectrum by considering the BPS spectrum of the $\mathcal S_0$ theory in (\ref{hdgT2}) and noting that the chiral primary operators $\psi_{-1/2}^{++}|0\rangle_{\mathrm{NS}}$ and $\tilde\psi_{-1/2}^{++}|0\rangle_{\mathrm{NS}}$ do not survive the action of the $\mathbb Z_2$ orbifold, see eq. (\ref{casei}). The BPS spectrum of the untwisted sector of the $\mathcal S_0/\mathbb Z_2$ theory will then be of the form $h_{0,0}=1$, $h_{1,0}=0$, $h_{0,1}=0$, and $h_{1,1}=1$. Since $\mathrm{S}^1$ has two fixed points under the action of the $\mathbb Z_2$, the twisted sector will contribute two more BPS states, leading to $h_{1/2,1/2}=2$. We then recover the Hodge-diamond (\ref{S0Z2hd}). 

Similarly, we can determine the reduced spectrum of the $\mathcal S_0/\mathbb Z_2$ theories with $(\mathcal N,\widetilde{\mathcal N})=(3,1)$, $(1,3)$ and $(1,1)$ supersymmetry. By this, we mean the orbifold even part of the BPS spectrum of $\mathcal{S}_0$. While not being chiral primary (since ${\cal N}=1$ supersymmetry does not possess chiral primary states), we still expect those states to be protected, because of the underlying \l SCA. The chiral primary $\psi_{-1/2}^{++}|0\rangle_{\mathrm{NS}}$ survives the action of the $\mathbb Z_2$ in case ($ii$) and so does its right-moving counterpart, see eq. (\ref{caseii}). Table \ref{hdgS0Z2untw} compares the spectra of the untwisted sectors of the $\mathcal S_0/\mathbb Z_2$ theories.
\begin{table}[h]
\begin{center}
    \begin{tabular}{ | c | c | c |}
    \hline
    {} & $\widetilde{\mathcal N}=3$ & $\widetilde{\mathcal N}=1$ \\ \hline
    $\mathcal N=3$ & 
  \begin{tabular}{ l c r }
  {} & 1 & {} \\
  0 & {} & 0 \\
  {} & 1 & {} \\
  \end{tabular} & 
  \begin{tabular}{ l c r }
  {} & 0 & {} \\
  0 & {} & 1 \\
  {} & 1 & {} \\
\end{tabular} \\ \hline
    $\mathcal N=1$ & 
    \begin{tabular}{ l c r }
  {} & 0 & {} \\
  1 & {} & 0 \\
  {} & 1 & {} \\
\end{tabular} & 
\begin{tabular}{ l c r }
  {} & 1 & {} \\
  1 & {} & 1 \\
  {} & 1 & {} \\
\end{tabular}  \\
    \hline
        \end{tabular}
\end{center}
\caption{Spectra of the untwisted sectors of the $\mathcal S_0/\mathbb Z_2$ theories. The $\mathbb Z_2$ group is acting such that it preserves either $\mathcal N=3$ (\ref{casei}) or $\mathcal N=1$ (\ref{caseii}) supersymmetry in the left and right-moving sectors.}
\label{hdgS0Z2untw}
\end{table}

An elliptic genus may then be derived for the $(\mathcal N,\widetilde{\mathcal N})=(3,3)$ theory and for the left and right-moving sectors of the (3,1) and (1,3) theories, respectively. We will discuss the elliptic genera in section \ref{CFT_ellgen}.

\subsection{BPS spectrum of symmetric orbifold of $\mathcal S_0/\mathbb Z_2$}\label{CFT_bps_symS0Z2}
We shall now compute the single-particle BPS spectrum of $\mathrm{Sym}^N(\mathcal{S}_0/\mathbb{Z}_2)$ with $\mathcal N=(3,3)$ supersymmetry. This is the quantity that we will compare and match with that of the stringy world-sheet theory as a first test of our proposal, see section \ref{sheet_bps_comp}. Since the fermionic boundary conditions are different between the odd and even twisted sectors, we treat each case separately. We follow closely the approach of \cite{Eberhardt:2017pty} and refer the reader to section 3 and appendix D of this paper for more details. The single-particle spectrum corresponds to the single-cycle states of the symmetric orbifold theory. We note that since the number of bosons and fermions are not equal in the seed theory, one has to be very careful when applying the DMVV formula \cite{Dijkgraaf:1996xw}. For this reason, we will derive the partition function and the associated BPS spectrum microscopically (see also the discussion at the end of section \ref{CFT_ellgen_sym}).

The seed theory $\mathcal S_0/\mathbb Z_2$ consists of the untwisted sector and twisted sector of $\mathbb Z_2$. Below we shall be explicit about the $\mathbb Z_2$ (un)twisted sectors and the symmetric orbifold odd and even twisted sectors to avoid confusions. We present the main results of the computations here and refer the reader to appendix \ref{app_SymN_bps} for the details.

\subsubsection{Odd twisted sector of symmetric orbifold}\label{CFT_bps_symS0Z2_odd}
\textbf{${\bf{\mathbb Z_2}}$ untwisted sector:}

We consider the theory in the NS sector. The seed theory is in the $\mathbb Z_2$ untwisted sector and has zero ground-state energy, see eq. (\ref{S0Z2_bps}). So the non-trivial contribution to the ground-state energy comes from the odd twisted sector of the symmetric orbifold, $n\in\mathbb Z_{\mathrm{odd>0}}$, were bosons and fermions have the same energy \cite{Lunin:2001pw}. Acting on the ground state with fractionally-moded fermionic fields, see eq. (\ref{SymNS0Z2oddn_i}), we find that there are two BPS states with conformal dimensions and $\mathfrak u(1)$-charges
\be\label{SymNS0UZ2oddn}
h=\bar{h}=\frac\ell2=\frac{\bar\ell}2=\frac{n\pm1}{4}\ .
\ee

\hspace{-20pt}
\textbf{${\bf{\mathbb Z_2}}$ twisted sector:}

The bosonic and fermionic fields have non-trivial ground-state energies in the $\mathbb{Z}_2$ twisted sector of the seed CFT. Taking this into account, we find that the associated BPS states have dimensions and charges
\be\label{SymNS0TZ2oddn}
h=\bar{h}=\frac\ell2=\frac{\bar\ell}2=\frac{n}{4}\ .
\ee
Moreover, this state appears with multiplicity 2.

\subsubsection{even twisted sector of symmetric orbifold}\label{CFT_bps_symS0Z2_even}
\textbf{${\bf{\mathbb Z_2}}$ untwisted sector:}

The even twisted sector of the symmetric orbifold is a bit more delicate. We start again with the $\mathbb Z_2$ untwisted sector of the seed CFT which has a vanishing ground-state energy. In the even twisted sector of the symmetric orbifold, fermions have a different boundary condition than the bosons and their ground-state energies are consequently different \cite{Lunin:2001pw}. We find that, applying fractionally-moded fermions, one cannot construct any BPS states in this sector as the conformal dimension of the ground-state is already high to begin with.

\vspace{10pt}\hspace{-20pt}
\textbf{${\bf{\mathbb Z_2}}$ twisted sector:}

Let us now describe the $\mathbb{Z}_2$-twisted sector. As discussed before, 3 of the 4 NS fermions in the seed theory are orbifolded by the $\mathbb Z_2$: this affects their boundary conditions and yields integer R-moded fermions. The one remaining fermion is not orbifolded and has half-integer NS modes. In the even twisted sector of the symmetric orbifold, the fermionic boundary conditions are switched again, but this time all the 4 fermions are acted on by the symmetric group. The 3 aforementioned fermions now become fractionally NS-moded and the remaining one becomes fractionally R-moded. The zero-modes of the R-moded fermion generate two states, but only one of them is orbifold even. Thus, if we find a BPS state, it will come with multiplicity 2 which is the multiplicity of the $\mathbb{Z}_2$ twisted sector. We find that one can indeed construct BPS states in this sector\footnote{This is not the case in the $\mathrm{Sym}^N(\mathcal S_0)$ theory: there are no BPS states in the even twisted sector of the symmetric orbifold, see \cite[section 3.4]{Eberhardt:2017pty}.} with dimensions and R-charges
\be\label{SymNS0TZ2evenn}
h=\bar{h}=\frac\ell2=\frac{\bar\ell}2=\frac{n}{4}\ .
\ee

\subsubsection{Full BPS spectrum}\label{CFT_bps_full}
All in all, taking the contributions from the odd and even twisted sectors of the symmetric orbifold together, we finally obtain the full BPS spectrum of $\mathrm{Sym}^N(\mathcal S_0/\mathbb Z_2)$:
\be\label{SymNS0Z2_bps}
\begin{tabular}{ccc}
&\vdots & \\
0 & & 0 \\
& 2 & \\
0 & & 0 \\
& 2 & \\
0 & & 0 \\
& 1 &
\end{tabular}
\qquad\oplus\quad
\begin{tabular}{ccc}
&\vdots & \\
 & 2&  \\
& 0 & \\
 &2 &  \\
& 0 & \\
 &2 &  \\
& 0 &
\end{tabular}
\qquad\oplus\quad
\begin{tabular}{ccc}
 & \vdots &  \\
& 0 & \\
& 2 & \\
& 0 & \\
& 2 & \\
& 0 & \\
& 0 &
\end{tabular}
\;=\quad
\begin{tabular}{ccc}
\vdots & \vdots & \vdots \\
0& 2 &0 \\
& 4 & \\
0& 2 &0 \\
& 4 & \\
0& 2 &0 \\
& 1 &
\end{tabular}\ ,
\ee
where the first, second and third terms on the LHS correspond to eq.s (\ref{SymNS0UZ2oddn}), (\ref{SymNS0TZ2oddn}), and (\ref{SymNS0TZ2evenn}), respectively. Eq. (\ref{SymNS0Z2_bps}) is the main result of this section. We will compare and match the CFT BPS spectrum to that of string theory in section \ref{sheet_bps}. 

In the conjectured dual CFT \eqref{number of copies}, there are $Q_1Q_5$ copies of the seed theory, so the maximal twist is $Q_1Q_5$. This sector has a BPS state with conformal weight $\tfrac{1}{4}Q_1Q_5=\tfrac{c}{12}$, where $c$ is the Brown-Henneaux central charge \eqref{Brown Henneaux central charge}. This is the stringy exclusion principle \cite{Maldacena:1998bw}.

\subsection{Moduli}\label{CFT_bps_moduli}
Having derived the BPS spectrum of $\mathrm{Sym}^N(\mathcal S_0/\mathbb Z_2)$, we shall now determine the moduli, i.e. exactly marginal operators of the theory. The moduli have dimensions $(h,\bar h)=(1,1)$ and are singlets of the R-symmetry in the left and right-moving sectors. The moduli are constructed by acting on the BPS states with $(h,\bar h)=(\tfrac12,\tfrac12)$ with the ${\cal N}=3$ supercharges $G^i_{-1/2}$. The $\mathrm{Sym}^N(\mathcal S_0/\mathbb Z_2)$ theory has four moduli, corresponding to the $h_{1,1}=4$ component of the Hodge diamond on the RHS of eq. (\ref{SymNS0Z2_bps}).

One modulus is in the untwisted sector of the $\mathbb Z_2$-orbifold of the seed theory as well as the untwisted sector of the symmetric orbifold and is the trivial modulus which changes the radius of S$^1$. Another modulus belongs to the untwisted sector of the $\mathbb Z_2$ and twist-3 sector of the symmetric orbifold. These two moduli are associated with the two moduli of the $\mathrm{Sym}^N(\mathcal S_0)$ theory and survive the $\mathbb Z_2$ action \cite{Gukov:2004ym}. The twist-3 sector modulus of $\mathrm{Sym}^N(\mathcal S_0)$ is identified with the RR axion in supergravity, see section 5.2 of this reference.

The interesting observation is that the $\mathrm{Sym}^N(\mathcal S_0/\mathbb Z_2)$ theory has two additional moduli which do not exist in $\mathrm{Sym}^N(\mathcal S_0)$. These moduli come from the $\mathbb Z_2$ twisted sector of the seed theory and from the twist-2 sector of the symmetric orbifold. It would be very interesting to understand the properties of the  dual moduli in supergravity.\footnote{We thank Matthias Gaberdiel for a discussion on this point.} We shall pursue this question in the near future.

Finally, we note that the exact marginality of these four operators (i.e., that they are indeed true moduli of the theory), is deduced by noting that these four operators are descendants of the $\mathcal N=(2,2)$ chiral primaries, see \cite{Dixon:1987bg} and \cite[appendix A]{Gukov:2004ym}.

\section{Elliptic genus of the dual CFT}\label{CFT_ellgen}
We start by computing the elliptic genus of the orbifold theory $\mathcal S_0/\mathbb Z_2$ in subsection \ref{CFT_ellgen_0}. We find that the elliptic genus for this theory vanishes due to the presence of a fermionic zero mode which survives the action of the orbifold. This is indeed the case for all theories which are $\mathcal N=3$ superconformal symmetric.

In subsection \ref{CFT_ellgen_mod} we consider the NS sector and determine a quantity which is composed of chiral primaries in the right-moving sector and of arbitrary excitations in the left-moving sector. This implies that the right-moving sector only consists of short representations whereas in the left-moving sector, both short and long representations contribute. The constraint of having half-integer spins in the NS sector yields $h-\bar h\in\mathbb Z/2$. Thus, the conformal dimensions on the left-moving sector have to be rigid, \emph{i.e.} do not acquire an anomalous dimension perturbatively. The quantity defined as such is an index and we refer to it as the ``modified" elliptic genus.

Finally, we compute the modified elliptic genus of the symmetric product orbifold of $\mathcal S_0/\mathbb Z_2$ in subsection \ref{CFT_ellgen_sym}. As we will show, the odd and even twisted sectors of the symmetric orbifold have to be examined separately and with care due to the presence of the fermionic zero mode in the seed theory $\mathcal S_0/\mathbb Z_2$.

We note that the modified elliptic genus is not modular invariant on its own: to obtain a modular invariant quantity one has to sum over the partition functions in the NS sector, with and without the insertion of $(-1)^{\rm F}$, as well as the partition function in the Ramond sector without the insertion of $(-1)^{\rm F}$, see footnote \ref{fn_mod_invar}. Nonetheless, as discussed above, it is an index and is invariant under deformations. We compare the modified genus of the proposed dual CFT to that of the string theory in section \ref{sheet_ellgen} and find that the two quantities match.

\subsection{Elliptic genus of $\mathcal S_0/\mathbb Z_2$}\label{CFT_ellgen_0}
The elliptic genus of an $\mathcal N=(2,2)$ CFT is defined as the trace over the Ramond-Ramond (RR) sector of the Hilbert space of the theory \cite{Witten:1993jg}:
\be\label{ellgen_dfnn}
\tilde{Z}_{\mathrm{R}}(z,\tau;0,\bar\tau)=\mathrm{tr}_{\mathrm{RR}}(-1)^{\rm F}q^{L_0-\frac{c}{24}}y^{J_0}\bar q^{\bar L_0-\frac{c}{24}}\ ,
\ee
where $J$ is the $\mathfrak{u}(1)$ R-current, $q=e^{2\pi i\tau}$, $y=e^{2\pi i z}$, and $(-1)^{\rm F}=(-1)^{{\rm F}_L}(-1)^{{\rm F}_R}$ with ${\rm F}_L$ and ${\rm F}_R$ being the left and right-moving fermion numbers, respectively. The tilde in $\tilde Z$ denotes the $(-1)^{\mathrm{F}}$ insertions in our conventions.

The partition function $\tilde{Z}_{\mathrm{R}}(z,\tau;\bar z,\bar\tau)$ can be computed similarly as \eqref{Z_NS_S0Z2}, and reads:
\begin{align}\label{Z_Rtilde_S0Z2}
\tilde{Z}_{\mathrm{R}}(z,\tau;\bar z,\bar\tau)&=\frac{1}{2}\Theta(\tau;\bar\tau;R)\left|\frac{\vartheta_1(z|\tau)\vartheta_1(\tau)}{\eta(\tau)^3}\right|^2+\left|\frac{\vartheta_2(z|\tau)\vartheta_1(\tau)^{\frac{1}{2}}\vartheta_2(\tau)^{\frac{1}{2}}}{\vartheta_2(\tau)^{\frac{1}{2}}\eta(\tau)^{\frac{3}{2}}}\right|^2\nonumber\\
&\!\!+\left|\frac{\vartheta_4(z|\tau)\vartheta_1(\tau)^{\frac{1}{2}}\vartheta_4(\tau)^{\frac{1}{2}}}{\vartheta_4(\tau)^{\frac{1}{2}}\eta(\tau)^{\frac{3}{2}}}\right|^2+\left|\frac{\vartheta_3(z|\tau)\vartheta_1(\tau)^{\frac{1}{2}}\vartheta_3(\tau)^{\frac{1}{2}}}{\vartheta_3(\tau)^{\frac{1}{2}}\eta(\tau)^{\frac{3}{2}}}\right|^2=0\ .
\end{align}
This function vanishes because of the existence of the fermionic zero-mode of the unorbifolded free fermion in the $\mathcal{N}=3$ algebra, see eq. (\ref{casei}). This is manifested in the presence of the $|\vartheta_1(\tau)^{\frac{1}{2}}|^2$ factor in all four terms of (\ref{Z_Rtilde_S0Z2}). The vanishing of the elliptic genus holds true for any CFT with $\mathcal{N}=3$ supersymmetry. This is because the linear $\mathcal N=3$ SCA contains a free fermion which is a singlet of the R-symmetry $\mathfrak{su}(2)_k$, see appendix \ref{app_reps_N3} and references \cite{Goddard:1988wv,Miki:1989ri}.\footnote{Analogously, the elliptic genera of any \l CFT with $\mathcal A_\gamma$ symmetry vanish.}

\subsection{Modified elliptic genus of $\mathcal S_0/\mathbb Z_2$}\label{CFT_ellgen_mod}
Typically, in CFTs with $\mathcal N=2$ or higher supersymmetry where all fermions are charged under the $\mathrm{U}(1)$ symmetry of the SCA, one can define four equivalent definitions of the elliptic genus through spectral flowing between the NS and R sectors and inserting (or not) the fermion number operators:\footnote{Note that even when not all of the fermions are charged under the $\mathfrak{u}(1)$, the first definition $\tilde Z_{\mathrm{R}}$ in eq. (\ref{ellgen_RtR}) is modular invariant while the other three transform into each other under modular transformations with their sum being a modular invariant quantity.\label{fn_mod_invar}}
\begin{align}
\mathcal{Z}(z,\tau)&\equiv \tilde{Z}_{\mathrm{R}}(z,\tau;\bar{z}=0,\bar{\tau})=Z_{\mathrm{R}}(z-\tfrac{1}{2},\tau;\bar{z}=-\tfrac{1}{2},\bar{\tau})\label{ellgen_RtR}\\
&=y^{-\frac{c}{6}} (q\bar{q})^{\frac{c}{24}}\tilde Z_{\mathrm{NS}}(z-\tfrac{\tau}{2},\tau;\bar{z}=-\tfrac{\bar{\tau}}{2},\bar{\tau})\label{ellgen_NSt}\\
&=y^{-\frac{c}{6}} (q\bar{q})^{\frac{c}{24}}{Z}_{\mathrm{NS}}(z-\tfrac{(\tau+1)}{2},\tau;\bar{z}=-\tfrac{(\bar{\tau}+1)}{2},\bar{\tau})\ .\label{ellgen_NS}
\end{align}
Here, $c$ is the central charge of the theory. This is because spectral flow interpolates between NS- and R-sectors for charged fermions. For the $\mathcal S_0/\mathbb Z_2$ theory, however, the four definitions in eqs.~(\ref{ellgen_RtR})--(\ref{ellgen_NS}) are not equal due to the presence of the uncharged fermion.

To define a non-vanishing index in a theory with charged fermionic zero-modes, one can take derivatives with respect to the chemical potential to obtain a non-vanishing protected quantity \cite{Gukov:2004fh, Maldacena:1999bp, Cecotti:1992qh}. Since in this case the fermionic zero-mode is uncharged, we cannot eliminate it in this way. Instead we will simply eliminate it by not using the definition of $\mathcal{Z}$ in the $\widetilde{\mathrm{R}}$-sector, but in another sector where the partition function does not vanish identically. Recall that in this case, spectral flow does not interpolate between the different sectors because uncharged fermions are unaffected by it and hence we cannot give a purely $\widetilde{\mathrm{R}}$-sector definition of our index.

The NS-sector definition with the insertion of $(-1)^{\rm F}$, $\tilde Z_{\textrm{NS}}$, is of interest for the purpose of comparison with the string theory and supergravity elliptic genus, as done for the cases of $\mathcal N=(2,2)$ and small $(4,4)$ AdS$_3$/CFT$_2$ dualities \cite{Datta:2017ert,deBoer:1998us}. We thus consider $\tilde Z_{\textrm{NS}}$, which defines a quantity composed of the contribution from chiral primaries in the right-moving sector and arbitrary excited states in the left-moving sector. This is because non-chiral primaries come always in pairs related by the action of the supercharge $\tilde{G}_{-1/2}$ resp.~$\tilde{G}_{-1/2}^+$ in the ${\cal N}=2$ and ${\cal N}=3$ cases. The two states cancel in the elliptic genus. Therefore, while only short representations contribute to the right-moving sector, both short and long representations contribute to the left-moving sector.

The conformal dimensions of the BPS states in the $\mathcal N=3$ algebra are quarter-integer: $h,\bar h\in\mathbb Z/4$, see eq.~(\ref{N3 BPS bound}). The requirement that the spins of the physical states are half-integer in the NS sector imposes the constraint $h-\bar h\in\mathbb Z/2$. This implies that the left-moving conformal dimensions, which may potentially come from long representations, have to be rigid and do not acquire perturbative corrections. We note that two short multiplets on the right may acquire corrections and join each other to form a long multiplet. For this to happen, the short multiplets have to come in a pair which cancels in the elliptic genus. Hence the elliptic genus remains invariant when short multiplets on the right lift. We thus conclude that the defined quantities in \eqref{ellgen_RtR}--\eqref{ellgen_NS} are indices.

We consider the $\widetilde{\rm NS}$ version of the elliptic genus:\footnote{For convenience, we have redefined the chemical potential $z$ in comparison to \eqref{ellgen_NSt}.}
\be\label{Z_NSt_S0Z2_i}
\tilde{\mathcal{Z}}_{\mathrm{NS}}(z,\tau)\equiv(q\bar{q})^{\frac{c}{24}}\tilde Z_{\mathrm{NS}}(z,\tau;\bar{z}=-\tfrac{\bar{\tau}}{2},\bar{\tau})=(q\bar{q})^{\frac{c}{24}}\tilde Z_{\mathrm{NS}}(z,\tau;\bar{z}=-\tfrac{\bar{\tau}}{2},\bar\tau)\Big|_{\bar h=\frac{\bar\ell}{2}}\ .
\ee
We call this index the ``modified" elliptic genus of the $\mathcal N=3$ CFT. The partition function $\tilde Z_{\mathrm{NS}}$ is again computed similar to \eqref{Z_NS_S0Z2} and is given by:
\begin{align} 
\tilde{Z}_{\mathrm{NS}}(z,\tau;\bar z,\bar\tau)&=\frac{1}{2}\Theta(\tau;\bar\tau;R)\left|\frac{\vartheta_4(z|\tau)\vartheta_4(\tau)}{\eta(\tau)^3}\right|^2+\left|\frac{\vartheta_3(z|\tau)\vartheta_4(\tau)^{\frac{1}{2}}\vartheta_3(\tau)^{\frac{1}{2}}}{\vartheta_2(\tau)^{\frac{1}{2}}\eta(\tau)^{\frac{3}{2}}}\right|^2+\nonumber\\
&\!\!+\left|\frac{\vartheta_1(z|\tau)\vartheta_4(\tau)^{\frac{1}{2}}\vartheta_1(\tau)^{\frac{1}{2}}}{\vartheta_4(\tau)^{\frac{1}{2}}\eta(\tau)^{\frac{3}{2}}}\right|^2+\left|\frac{\vartheta_2(z|\tau)\vartheta_4(\tau)^{\frac{1}{2}}\vartheta_2(\tau)^{\frac{1}{2}}}{\vartheta_3(\tau)^{\frac{1}{2}}\eta(\tau)^{\frac{3}{2}}}\right|^2\label{Z_NSt_S0Z2_ii}\\
&\!\!=\frac{1}{2}\Theta(\tau)\left|\frac{\vartheta_4(z|\tau)\vartheta_4(\tau)}{\eta(\tau)^3}\right|^2+2\left|\frac{\vartheta_3(z|\tau)}{\vartheta_2(\tau)}\right|^2+2\left|\frac{\vartheta_2(z|\tau)}{\vartheta_3(\tau)}\right|^2\ .\label{Z_NSt_S0Z2_iii}
\end{align}
This is a non-vanishing quantity at $\bar z=-\tfrac{\bar\tau}{2}$. The two terms in the first line of eq. (\ref{Z_NSt_S0Z2_ii}) are contributions from the untwisted sector of the $\mathbb Z_2$ orbifold and the two terms in the second line come from the $\mathbb Z_2$ twisted sector, see eq.s (\ref{unt_unp})-(\ref{Z_NS_S0Z2}). The modified elliptic genus (\ref{Z_NSt_S0Z2_i}) then reads
\be 
\tilde{\mathcal{Z}}_{\mathrm{NS}}(z,\tau)=2q^{\frac{1}{8}}\left(\frac{\vartheta_3(z|\tau)}{\vartheta_2(\tau)}+\frac{\vartheta_2(z|\tau)}{\vartheta_3(\tau)}\right)\ .
\ee
Note that the result is independent of $\bar{q}$.

\subsection{Modified elliptic genus of symmetric orbifold of $\mathcal S_0/\mathbb Z_2$}\label{CFT_ellgen_sym}
Having defined the modified elliptic genus of $\mathcal S_0/\mathbb Z_2$, we shall now compute the modified elliptic genus of its symmetric product orbifold. As discussed in the previous subsection, the modified elliptic genus of the seed theory is an index and so is the modified elliptic genus of the symmetric orbifold of it. One would then expect that, in the context of the proposed $\mathcal N=(3,3)$ duality, the modified elliptic genus matches its counterpart computed in string theory. We will verify this in section \ref{sheet_ellgen}.

In the comparison between the modified elliptic genus of the dual CFT and that of the supergravity (i.e. low energy) limit of string theory on the $\mathrm{AdS}_3\times(\mathrm{S}^3\times\mathrm{S}^3\times\mathrm{S}^1)/\mathbb Z_2$ background, we will focus on the contribution from single-particle states in the spectrum. These correspond to single-cycle states of the symmetric orbifold. Multi-particle states correspond to multiple cycles.

In the remainder of this section we compute the modified elliptic genus of the symmetric orbifold of $\mathcal S_0/\mathbb Z_2$ from first principles. We note that there are two orbifold actions at work: the first one is the $\mathbb Z_2$ action on $\mathcal S_0$ which yields an untwisted as well as a twisted sector to the Hilbert space of the seed theory. The second orbifold action is that of the symmetric product of $N=Q_1Q_5$ copies of the seed theory, $\mathrm{Sym}^N(\mathcal S_0/\mathbb Z_2)$, and contributes to twist-$n$ sectors with $1\le n\le N$.\footnote{One could view this also as an orbifold of $\mathcal{S}_0^N$ by the wreath product of the symmetric group with $\mathbb{Z}_2$, we prefer however to view the theory as an orbifold of an orbifold.} We shall be explicit below about the $\mathbb Z_2$ versus $\mathrm{Sym}^N$ orbifold actions to avoid confusions. We present the details of the analysis of the odd twisted sector in appendix \ref{app_SymN_ellgen}.

\subsubsection{Odd twisted sector of symmetric orbifold}\label{CFT_ellgen_sym_odd}
We start by considering the odd twisted sector. The contribution of single-particle states comes from states comprised of one cyclic permutation of $n$ copies of the seed theory ($n\in\mathbb Z_{\mathrm{odd}}$, $1\le n\le N$) with symmetrised excitations on the $n$ copies, and $N-n$ copies of the seed theory in their vacua. The untwisted projected partition function (in either NS or R sector) reads
\be\label{Z_twistodd}
Z(z,\tau)=Z(nz,n\tau)\big(Z(z,\tau)\big)^{N-n}\ .
\ee

We next perform an S-modular transformation to obtain the partition function of the twist-$n$ sector in terms of the fractional modes of bosons and fermions in the twist $n$ sector, see \cite[appendix D]{Eberhardt:2017pty} and \cite[appendix A]{Gaberdiel:2018rqv}. This operation  depends on the choice of the sector. For the NS sector and without $(-1)^{\rm F}$ insertions, we have
\be\label{Z_twistodd_ii}
Z_{\mathrm{NS}}(z,\tau)=Z_{\mathrm{NS}}(z,\tfrac{\tau}n)\big(Z_{\mathrm{NS}}(z,\tau)\big)^{N-n}\ .
\ee
This is because under an S-modular transformation, the NS sector partition function is mapped to itself. However, $\tilde Z_{\mathrm{NS}}(nz,n\tau)$ (with fermion number insertions) is transformed into $Z_{\mathrm{R}}(z,\tfrac{\tau}n)$. Similarly, $\tilde Z_{\mathrm{R}}$ transforms into itself whereas $Z_{\mathrm{R}}(nz,n\tau)$ is transformed into $\tilde Z_{\mathrm{NS}}(z,\tfrac{\tau}n)$. Since we are interested in computing $\tilde Z_{\mathrm{NS}}(z,\frac{\tau}n)$ for our modified elliptic genus (\ref{Z_NSt_S0Z2_i}), we need to start from the partition function $Z_{\mathrm{R}}(nz,n\tau)$ in (\ref{Z_twistodd}) and then perform the S-modular transformation to obtain $\tilde Z_{\mathrm{NS}}(z,\tfrac{\tau}n)$.

The full derivation of the modified elliptic genus is presented in appendix \ref{app_SymN_ellgen}. Here we write the final result:
\bea\label{ellgen_NSt_oddSym}
\!\!\!\!\!\tilde{\mathcal{Z}}_{\mathrm{NS}}(z,\tau)=2\sum_{n\text{ odd}}
\bigg(q^{\frac n4}\,y^{\frac n2}\,\frac{\vartheta_2(z+\tfrac{\tau}2|\tfrac\tau n)}{\vartheta_2(\tfrac\tau n)}\Bigg|_{h\in \frac{\mathbb Z}{2}}+
q^{\frac n4}\,y^{\frac n2}\,\frac{\vartheta_3(z+\tfrac{\tau}2|\tfrac\tau n)}{\vartheta_3(\tfrac\tau n)}\Bigg|_{h\in \frac{\mathbb Z}{2}+\frac{1}{4}}\bigg)\ .
\eea

We would next like to Fourier expand this result for the purpose of comparing it with our worldsheet string theory analyses in section \ref{sheet_ellgen}. To do so, we find it more convenient to analyse each term in equation (\ref{ellgen_NSt_oddSym}) separately. The first term on the RHS comes from the $\mathbb Z_2$ untwisted sector whereas the second term is the $\mathbb Z_2$ twisted sector contribution. We define:
\bea
&&\tilde{\mathcal{Z}}_{\mathrm{NS}}^{\mathrm{U_{\mathbb Z_2}}}(z,\tau)=2 \sum_{n\text{ odd}} 
q^{\frac n4}\,y^{\frac n2}\,\frac{\vartheta_2(z+\tfrac{\tau}2|\tfrac\tau n)}{\vartheta_2(\tfrac\tau n)}\Bigg|_{h\in \frac{\mathbb Z}{2}}\ ,\label{ellgen_NSt_sym_oddt_iv}\\
&&\tilde{\mathcal{Z}}_{\mathrm{NS}}^{\mathrm{T_{\mathbb Z_2}}}(z,\tau)=2 \sum_{n\text{ odd}} 
q^{\frac n4}\,y^{\frac n2}\,\frac{\vartheta_3(z+\tfrac{\tau}2|\tfrac\tau n)}{\vartheta_3(\tfrac\tau n)}\Bigg|_{h\in \frac{\mathbb Z}{2}+\frac{1}{4}}\ ,\label{ellgen_NSt_sym_oddt_v}
\eea
where the superscripts $\mathrm{U_{\mathbb Z_2}}$ and $\mathrm{T_{\mathbb Z_2}}$ correspond to the $\mathbb Z_2$ untwisted and twisted sectors, respectively.

The details of the computation are again presented in appendix \ref{app_SymN_ellgen}. Using the quasi-periodicity properties of the Fourier coefficients of the modified elliptic genus, we find that:

\bea\label{ellgen_NSt_oddSym_UZ2}
&&\tilde{\mathcal{Z}}_{\mathrm{NS}}^{\mathrm{U_{\mathbb Z_2}}}(z,\tfrac\tau n)
=\sum_{m^\prime\in\frac{\mathbb Z}2,\,\ell^\prime\in\mathbb Z}
\Big(2\delta_{m^\prime,\pm\frac{\ell^\prime}2}\delta_{m^\prime,\frac12\mathbb Z_{>0}}-3\delta_{m^\prime,0}\delta_{\ell^\prime,0}\Big)\,q^{m^\prime}y^{\ell^\prime}\nonumber\\
&&\qquad\qquad\;\;=\frac{2}{1-yq^{\frac12}}+\frac{2}{1-y^{-1}q^{\frac12}}-3\ ,
\eea
and
\bea\label{ellgen_NSt_oddSym_TZ2}
&&\tilde{\mathcal{Z}}_{\mathrm{NS}}^{\mathrm{T_{\mathbb Z_2}}}(z,\tfrac\tau n)
=\sum_{m^\prime\in\frac{\mathbb Z}2+\frac14,\,\ell^\prime\in\mathbb Z+\frac12}
2\delta_{m^\prime,\pm\frac{l^\prime}2}\delta_{m^\prime,\frac12\mathbb Z_{\ge  0}+\frac14}\,q^{m^\prime}y^{l^\prime}\nonumber\\
&&\qquad\qquad\;\;=\frac{2y^{\frac12}q^{\frac14}}{1-yq^{\frac12}}+\frac{2y^{-\frac12}q^{\frac14}}{1-y^{-1}q^{\frac12}}\ .
\eea
We compare and match separately the modified elliptic genera (\ref{ellgen_NSt_oddSym_UZ2}) and (\ref{ellgen_NSt_oddSym_TZ2}) with their counterparts in the $\mathbb Z_2$ untwisted and twisted sectors string theory in section \ref{sheet_ellgen}.

\subsubsection{Even twisted sector of symmetric orbifold}\label{CFT_ellgen_sym_even}
We shall now turn to the computation of the modified elliptic genus in the even twisted sector of the symmetric orbifold theory. Fermionic fields acquire a minus sign under even cyclic permutations and have R-moding in the NS sector and vice versa. Following our procedure in the previous subsection, we thus need to determine the partition function of the symmetric orbifold $\tilde Z_{\mathrm{R}}(nz,n\tau)$, perform the S-modular transformation, and obtain the R-type partition function $\tilde Z_{\mathrm{R}}(z,\tfrac{\tau}n)$ which has the appropriate moding for the even twisted sector, see the discussion below eq. (\ref{Z_twistodd_ii}).

The R sector partition function $\tilde Z_{\mathrm{R}}$ of the seed theory was, however, computed earlier in (\ref{Z_Rtilde_S0Z2}) where we found that it vanishes identically due to the zero mode of the uncharged fermion in the \3 SCA. This continues to hold true in the symmetric product orbifold. We thus conclude that the modified elliptic genus in the even twisted sector of the symmetric orbifold vanishes.

Finally, we note that if we apply the DMVV formula only to the odd twisted sector of the symmetric orbifold, where bosons and fermions have the same ground-state energies, we would indeed obtain eq.s (\ref{ellgen_NSt_oddSym_UZ2}) and (\ref{ellgen_NSt_oddSym_TZ2}), as expected. This, however, is not the case for the even twisted sector.\footnote{We thank Christoph Keller for a discussion on this point.}

\section{BPS spectrum in string theory}\label{sheet_bps}
In this section, we analyse the string worldsheet theory. Since this is largely an amalgamation of \cite{Yamaguchi:1999gb}, \cite{Eberhardt:2017pty} and \cite{Datta:2017ert}, we will follow these sources closely.

We will heavily rely on the representation theory of $\mathfrak{osp}(3|2)$, the global subalgebra of the ${\cal N}=3$ SCA, as discussed in Appendix~\ref{app_alg}. The BPS condition for this algebra is given in \eqref{N3 BPS bound}, we reproduce it here for completeness:
\be
h_{\mathrm{BPS}}=\frac \ell2\ ,
\ee
where $\ell$ is the $\mathfrak{su}(2)$ spin and $\ell\le\textstyle\frac k2$, where $k$ is the level of the affine $\mathfrak{su}(2)_k$ algebra inside the ${\cal N}=3$ SCA.

The unorbifolded worldsheet theory with pure NS-NS flux is a WZW model based on
\be 
\mathfrak{sl}(2,\mathbb{R})^{(1)}_{\frac{k^+k^-}{k^++k^-}} \oplus \mathfrak{su}(2)^{(1)}_{k^+} \oplus \mathfrak{su}(2)^{(1)}_{k^-} \oplus \mathfrak{u}(1)^{(1)}\ .
\ee
Here, the superscript $(1)$ indicates that this is an $\mathcal{N}=1$ supersymmetric affine WZW model. For more details on these algebras, see e.g.~\cite{Gaberdiel:2013vva}. These algebras split into bosonic WZW models with level shifted by the respective dual Coxeter number of the algebra together with free fermions. Thus, the bosonic (or 'decoupled') currents generate the affine algebra
\be
\mathfrak{sl}(2,\mathbb{R})_{\frac{k^+k^-}{k^++k^-}+2} \oplus \mathfrak{su}(2)_{k^+-2} \oplus \mathfrak{su}(2)_{k^--2} \oplus \mathfrak{u}(1)\ .
\ee
From this, it is clear that we should require $k^\pm \ge 2$ in order for the worldsheet theory to be unitary.\footnote{See however \cite{Gaberdiel:2018rqv} for a recent proposal of how to make sense of the $k^\pm=1$ case.}
 We denote the $\mathfrak{sl}(2,\mathbb{R})$-fermions by $\psi^a$, $a \in \{3,\pm\}$ and the $\mathfrak{su}(2)^{\pm}$-fermions by $\chi^{\pm, a}$. The $\mathfrak{u}(1)$-fermion is denoted by $\lambda$. Since these fermions are almost all we need to construct BPS states, we will focus on those. 

To take the orbifold, we have to require that $k^+=k^-$. The level of the diagonal $\mathfrak{su}(2)$ is then $k=k^++k^-=2k^+$ and the level of the $\mathfrak{sl}(2,\mathbb{R})$-algebra is given by $\tfrac{1}{2}k^+=\tfrac{1}{4}k$.\footnote{Note that $k$ is always even. We chose the conventions in such a way that $k$ corresponds to the level of the $\mathcal{N}=3$ SCA as first given in \cite{Miki:1989ri} and reviewed in Appendix~\ref{app_alg}.} To distinguish the $\mathfrak{su}(2)$'s, we will denote them by $\mathfrak{su}(2)^+$, $\mathfrak{su}(2)^-$ and the diagonal $\mathfrak{su}(2)$ simply by $\mathfrak{su}(2)$. We will also follow the following convention for the spins: $j$ will always denote the $\mathfrak{sl}(2,\mathbb{R})_{k/4}$-spin, which corresponds to the conformal weight $h$ of the dual CFT. $\ell^+$ and $\ell^+$ denote the spins of $\mathfrak{su}(2)^+_{k/2}$ and $\mathfrak{su}(2)^-_{k/2}$. Finally, $\ell$ denotes the spin of $\mathfrak{su}(2)_k$.

\subsection{Review of the unorbifolded theory}\label{sheet_bps_review}
We start by reviewing the unorbifolded theory and its BPS spectrum. This was worked out in \cite{Eberhardt:2017fsi}. We will focus in the following on the massless field content. These are fields with the minimal excitation level, which is $\tfrac{1}{2}$ in the NS sector and $0$ in the R sector. Let us explain why these fields correspond precisely to the supergravity KK-modes. The supergravity limit can by obtained by the limit $k \to \infty$. In this case, the string tension compared to the $\mathrm{AdS}_3$-radius becomes very large and stringy excitations become very heavy. The corresponding three-dimensional mass can be computed via the relation $m^2=(j+\bar{j})(j+\bar{j}-2)/k$ \cite{Aharony:1999ti, deBoer:1998kjm}. It is indeed true that except for the massless excitations, the mass becomes infinite in the limit $k \to \infty$.\footnote{There is a small subtlety in this statement: when including the so-called spectrally flowed sectors of the affine $\mathfrak{sl}(2,\mathbb{R})_{k/4}$-algebra, we also have to use oscillator excitations. For simplicity's sake, we assume in this paper that $k$ is large and do not consider spectrally flowed sectors. We have however checked that the arguments we present below also go through in the spectrally flowed case.}

In \cite{Eberhardt:2017fsi}, it was shown that indeed all BPS states of string theory on $\mathrm{AdS}_3 \times \mathrm{S}^3\times \mathrm{S}^3 \times \mathrm{S}^1$ come from massless fields. This was to be expected, since these are the only fields surviving the $k \to \infty$ limit. Since BPS states should be protected, we expect them to be among the massless fields. For this reason, we will in the following only discuss massless fields.

The partition function of string theory on $\mathrm{AdS}_3 \times \mathrm{S}^3 \times \mathrm{S}^3 \times \mathrm{S}^1$ involves a sum over the $\mathfrak{su}(2)^\pm$-spins. It was demonstrated in \cite{Eberhardt:2017fsi} that we obtain BPS states only in the case $\ell^+=\ell^-$, i.e.~when both spins agree. In this case, there is for every spin one BPS state coming from the NS and one from the R sector. When combining left- and right-movers, this yields one BPS state from the NS-NS, the NS-R, the R-NS and the R-R sector, respectively. In the NS sector, this BPS state has the form $\psi^-_{-1/2} |j_0=\ell_0^++1,\ell_0^+,\ell_0^+\rangle$. The three quantum numbers label the $\mathfrak{sl}(2,\mathbb{R})$- and the two $\mathfrak{su}(2)$-spins of the ground state. This state satisfies all physical state conditions: (\textit{i}) it is trivially annihilated by all positive $L_n$-modes, (\textit{ii}) it is annihilated by $G_{1/2}$,\footnote{This is the supercharge of the $\mathcal{N}=1$ worldsheet SCA.} (\textit{iii}) it satisfies the GSO-projection, and (\textit{iv}) it satisfies the mass-shell condition. It was checked in \cite{Eberhardt:2017fsi} that restricting to superprimary fields and eliminating null-fields can be done analogously to the light-cone gauge. It has simply the effect of removing two uncharged oscillators. The mass-shell condition for a general state is given by
\be 
-\frac{4j_0(j_0-1)}{k}+\frac{2\ell_0^+(\ell_0^++1)}{k}+\frac{2\ell_0^-(\ell_0^-+1)}{k}+N=a\ .
\ee
Here, $a$ is the normal-ordering constant, which equals $\tfrac{1}{2}$ in the NS sector and $0$ in the R sector. To restrict to massless fields, we set $N=a$. The quantum numbers $j_0$, $\ell_0^+$ and $\ell^-_0$ refer to the ground-state spins, whereas $j$, $\ell^+$ and $\ell^-$ are the actual spins. For the state $\psi^-_{-1/2} |j_0=\ell_0^++1,\ell_0^+,\ell_0^+\rangle$, the mass-shell condition is clearly satisfied and we have
\be 
j=j_0-1=\ell_0^+=\ell_0^-=\ell^+=\ell^-\ .
\ee
This is because we used one $\mathfrak{sl}(2,\mathbb{R})$-charged oscillator. Since $j$ is identified with $h$ in the dual CFT, this state saturates the $\mathcal{N}=4$ BPS bound \eqref{N4 BPS bound}.

In the R sector, the fermionic zero-modes generate an 8-dimensional representation of the algebra $\mathfrak{sl}(2,\mathbb{R})\oplus \mathfrak{su}(2) \oplus \mathfrak{su}(2)$, namely $(\mathbf{2},\mathbf{2},\mathbf{2})$. The eight states have spins $|j_0 \pm \tfrac{1}{2},\ell_0^+\pm \tfrac{1}{2},\ell_0^- \pm \tfrac{1}{2} \rangle$. Of this representation, we may pick the state $|j_0 - \tfrac{1}{2},\ell_0^++ \tfrac{1}{2},\ell_0^- + \tfrac{1}{2} \rangle$. Let us again restrict to the case $\ell^+_0=\ell^-_0$, otherwise this does not yield a BPS state. The state satisfies the mass-shell condition, provided that $j_0=\ell_0^++1=\ell_0^-+1$ and hence
\be 
j=j_0-\tfrac{1}{2}=\ell_0^++\tfrac{1}{2}=\ell_0^-+\tfrac{1}{2}=\ell^+=\ell^-\ .
\ee
So, this state is again a BPS state.

To summarize, the complete BPS spectrum of string theory on $\mathrm{AdS}_3 \times \mathrm{S}^3 \times \mathrm{S}^3 \times \mathrm{S}^1$ in the limit $k \to \infty$ is
\begin{align} 
\bigoplus_{\ell^+=\ell^-\in \frac{1}{2}\mathbb{Z}_{\ge 0}} &([\ell^+,\ell^-,u=0]_\mathrm{S} \oplus [\ell^++\tfrac{1}{2},\ell^-+\tfrac{1}{2},u=0]_\mathrm{S})\nonumber\\
&\qquad\otimes ([\ell^+,\ell^-,u=0]_\mathrm{S} \oplus [\ell^++\tfrac{1}{2},\ell^-+\tfrac{1}{2},u=0]_\mathrm{S})\ . \label{large N4 BPS spectrum}
\end{align}
Here, $[\ell^+,\ell^-,u]_\mathrm{S}$ refers to a short $\mathfrak{d}(2,1;\alpha)$, the global subalgebra of the large ${\cal N}=4$ SCA, multiplet. The structure and character of such a multiplet is discussed in Appendix~\ref{app_alg}. The first factor corresponds to the left-movers and the second factor to the right-movers. The first summand of each factor is the $\mathcal{N}=4$ multiplet with highest weight state in the NS sector, the second summand has its highest weight state in the R sector.

\subsection{The $\mathbb Z_2$ untwisted sector}\label{sheet_bps_Z2U}
We first discuss the untwisted sector of the $\mathbb{Z}_2$-orbifold. As explained above, we restrict to the unflowed sector of $\mathfrak{sl}(2,\mathbb{R})_{k/4}$. The $\mathbb{Z}_2$ acts as follows on the fermions:
\be 
(\psi,\chi^{+,a},\chi^{-,a},\lambda) \longmapsto (\psi,\chi^{-,a},\chi^{+,a},-\lambda)\ .
\ee

Let us first discuss the NS sector. The BPS state was reviewed above and has the form $\psi^-_{-1/2} |j_0=\ell_0^++1,\ell_0^+,\ell_0^+\rangle$. It is obviously invariant under the orbifold action, since the $\mathfrak{sl}(2,\mathbb{R})$-oscillators are orbifold-invariant.

Next, we treat the R sector. This is much more complicated. Here, the fermions have zero-modes, which generate the representation $2(\mathbf{2},\mathbf{2},\mathbf{2})$ of $\mathfrak{sl}(2,\mathbb{R}) \oplus \mathfrak{su}(2)^+ \oplus \mathfrak{su}(2)^-$. The factor $2$ is removed by the GSO-projection, but for now, we work with the fermions before the GSO projection. After orbifolding, this representation branches down to $2(\mathbf{2},\mathbf{1}) \oplus 2(\mathbf{2},\mathbf{3})$ of $\mathfrak{sl}(2,\mathbb{R}) \oplus \mathfrak{su}(2)$. It is however quite non-trivial to see how the orbifold and GSO projection act on these states.

We can fix the representation content of the R sector as follows. The NS sector fermionic partition function is (including physical state conditions):
\begin{align} 
&\frac{1}{2}\Bigg(\left|\frac{\vartheta_3(u|\tau)\vartheta_3(z|\tau)^2\vartheta_3(\tau)}{\eta(\tau)^4}-\frac{\vartheta_4(u|\tau)\vartheta_4(z|\tau)^2\vartheta_4(\tau)}{\eta(\tau)^4}\right|^2\nonumber\\
&\qquad+\left|\frac{\vartheta_3(u|\tau)\vartheta_3(z|\tau)\vartheta_4(z|\tau)\vartheta_4(\tau)}{\eta(\tau)^4}-\frac{\vartheta_4(u|\tau)\vartheta_4(z|\tau)\vartheta_3(z|\tau)\vartheta_3(\tau)}{\eta(\tau)^4}\right|^2\Bigg)\ .
\end{align}
Here, $u$ is the $\mathfrak{sl}(2,\mathbb{R})$-potential and $z$ is the $\mathfrak{su}(2)$-potential. The first line corresponds to the unprojected partition function. We have two fermions charged under $\mathfrak{sl}(2,\mathbb{R})$ (namely $\psi^\pm$), four fermions charged under $\mathfrak{su}(2)$ (namely $\chi^{\pm,\pm}$) and two uncharged fermions (e.g.~$\chi^{+,3}-\chi^{-,3}$ and $\lambda$, remember that two uncharged fermions are eliminated due to the physical state conditions). The second term in the first line takes care of the GSO-projection. The second line is the projected partition function. Two uncharged fermions are orbifolded, since both $\chi^{+,3}-\chi^{-,3}$ and $\lambda$ are odd.\footnote{We had to eliminate the two unorbifolded fermions in the light-cone gauge to be able to correctly orbifold the theory.} Also two of the $\mathfrak{su}(2)$-charged fermions are orbifolded, namely $\chi^{+,\pm}-\chi^{-,\pm}$. The other two fermions $\chi^{+,\pm}+\chi^{-,\pm}$ are invariant.

In the R sector, $\vartheta_3$'s become $\vartheta_2$'s and $\vartheta_4$'s become $\vartheta_1$'s. Thus, the second term in the first line and the first term in the second line yields a $\vartheta_1(\tau)=0$ in the R sector. Thus, the complete expression including the R sector is
\begin{align} 
&\frac{1}{2}\Bigg(\Bigg|\frac{\vartheta_3(y|\tau)\vartheta_3(z|\tau)^2\vartheta_3(\tau)}{\eta(\tau)^4}-\frac{\vartheta_4(y|\tau)\vartheta_4(z|\tau)^2\vartheta_4(\tau)}{\eta(\tau)^4}+\frac{\vartheta_2(y|\tau)\vartheta_2(z|\tau)^2\vartheta_2(\tau)}{\eta(\tau)^4}\Bigg|^2\nonumber\\
&\qquad+\Bigg|\frac{\vartheta_3(y|\tau)\vartheta_3(z|\tau)\vartheta_4(z|\tau)\vartheta_4(\tau)}{\eta(\tau)^4}-\frac{\vartheta_4(y|\tau)\vartheta_4(z|\tau)\vartheta_3(z|\tau)\vartheta_3(\tau)}{\eta(\tau)^4}\nonumber\\
&\qquad\qquad\qquad\qquad\qquad\qquad\qquad\qquad\pm\frac{\vartheta_1(y|\tau)\vartheta_1(z|\tau)\vartheta_2(z|\tau)\vartheta_2(\tau)}{\eta(\tau)^4} \Bigg|^2\Bigg)\ .
\end{align}
The sign of the last term is ambiguous and can be chosen independently for left- and right-movers. We will discuss the impact of the corresponding sign choice below. The zero-modes (i.e.~the coefficient of $q^{\frac{1}{2}}$, which corresponds to the massless states) read
\begin{align} 
&\frac{1}{2}\Big(\left|v+v^{-1}+2(y+y^{-1})+2+(v^{\frac{1}{2}}+v^{-\frac{1}{2}})(y^{\frac{1}{2}}+y^{-\frac{1}{2}})^2\right|^2\nonumber\\
&\qquad+\left|v+v^{-1}-2\pm (v^{\frac{1}{2}}-v^{-\frac{1}{2}})(y^{\frac{1}{2}}-y^{-\frac{1}{2}})(y^{\frac{1}{2}}+y^{-\frac{1}{2}})\right|^2 \Big)\\
=& \frac{1}{2}\Big(\left|v^{-1} (1+v^{\frac{1}{2}}y)(1+v^{\frac{1}{2}}y^{-1})(1+v^{\frac{1}{2}})^2\right|^2\nonumber\\
&\qquad+\left|v^{-1} (1\mp v^{\frac{1}{2}}y)(1\pm v^{\frac{1}{2}}y^{-1})(1+v^{\frac{1}{2}})(1-v^{\frac{1}{2}})\right|^2\Big)\ .
 \end{align}
where $v=\mathrm{e}^{2\pi\mathrm{i} u}$ and $y=\mathrm{e}^{2\pi\mathrm{i}z}$. 

Let us discuss this result. The first term is the unprojected partition function. As one can see, it factorizes into four factors, corresponding to the action of the four supercharges of the large ${\cal N}=4$ SCA transforming in the representation $\mathbf{3}\oplus \mathbf{1}$ of $\mathfrak{su}(2)$. Thus, the partition function of $\mathrm{AdS}_3 \times \mathrm{S}^3 \times \mathrm{S}^3 \times  \mathrm{S}^1$ is manifestly $\mathcal{N}=4$ supersymmetric. The second term show that some of the supercharges are orbifolded out --- they act trivially on the BPS states. The positively charged supercharge $G^+_{-1/2}$ is orbifolded out, provided that we choose the plus-sign for both the left- and right-movers. For the other sign-choice the negatively charged supercharge $G^-_{-1/2}$ is orbifolded out. The four possibilities on how to perform the orbifold correspond precisely to the $\mathcal{N}=(3,3)$, $\mathcal{N}=(3,1)$, $\mathcal{N}=(1,3)$ and the $\mathcal{N}=(1,1)$ theories discovered in \cite{Yamaguchi:1999gb}. The four possibilities also simply correspond to to the four possible superstring theories on the background, they are the four different GSO-projections. We usually only talk about two different GSO-projections, but for this background all four of them are inequivalent. For the background $\mathrm{AdS}_3 \times \mathrm{S}^3 \times \mathrm{S}^3 \times \mathrm{S}^1$, there was no sign-ambiguity, since we can T-dualize along the circle $\mathrm{S}^1$ to see that the type IIA and IIB superstrings are in fact equivalent.

From this partition function, we can also read off the action of the $\mathbb{Z}_2$ on the BPS states. The BPS states have to be of the form $(v^{\frac{1}{2}} u)^\ell(\bar{v}^{\frac{1}{2}} \bar{u})^{\bar{\ell}}$ by virtue of the BPS bound \eqref{N3 BPS bound}. We see that these are generated by the first of the four terms in the product. So the BPS partition function reads
\be 
\frac{1}{2}\Big(\left|1+v^{\frac{1}{2}}u\right|^2+\left|1\mp v^{\frac{1}{2}}u\right|^2\Big)\ .
\ee
This precisely reproduces Table~\ref{hdgS0Z2untw}. In the following we will restrict to the case where we have $\mathcal{N}=(3,3)$ supersymmetry.
 
To summarize, the supergravity BPS spectrum from the untwisted sector reads
\be 
\bigoplus_{\ell\in \mathbb{Z}_{\ge 0}}[\ell]_{\mathrm{S}} \otimes [\ell]_{\mathrm{S}}\oplus [\ell+1]_{\mathrm{S}} \otimes [\ell+1]_{\mathrm{S}}\ . \label{worldsheet_untwisted_BPS spectrum}
\ee
Note that the $\mathfrak{su}(2)$-spin only takes integer values, since $\ell=\ell^++\ell^-=2\ell^+ \in \mathbb{Z}$. $[\ell]_{\mathrm{S}}$ denotes a short $\mathcal{N}=3$ character.

\subsection{The $\mathbb Z_2$ twisted sector}\label{sheet_bps_Z2T}
In this subsection, we treat the $\mathbb{Z}_2$ twisted sector. 

Let us begin with the twisted NS sector. For this, note that the worldsheet theory is almost a symmetric product orbifold with two copies of $\mathcal{S}_{k/2-2}$. In particular, the ground-state energy of the twisted sector agrees with what was calculated in \cite[Equation (D.8)]{Eberhardt:2017pty} for the twist-2 sector of the symmetric product orbifold and reads
\be 
h=\frac{c}{12}+\frac{2}{8k}=\frac{2(k/2-1)}{2k}+\frac{2}{8k}=\frac{2k-3}{4k}\ .
\ee
The fermions $\chi^a$ are moded in $\tfrac{1}{2}\mathbb{Z}$ in the twisted sector\footnote{From the symmetric orbifold perspective there is only one set of $\chi$'s. We could equally well still use the combinations $\chi^{+,a}+\chi^{-,a}$ and $\chi^{+,a}-\chi^{-,a}$. The former would be NS-moded, while the latter combination is R-moded.} and the fermion $\lambda$ is R-moded. $\psi$ is still NS-moded. We have four chiral fermionic zero-modes (i.e.~we have also the four right-moving zero-modes): $\chi_0^a$ and $\lambda_0$. They transform in the representation $\mathbf{3} \oplus \mathbf{1}$ of $\mathfrak{su}(2)$, hence their zero-modes generate the $4=2^2$-dimensional representation $2 \cdot \mathbf{2}$ of $\mathfrak{su}(2)$. This is the associated spinor representation of $\mathbf{3} \oplus \mathbf{1}$. Now we need to impose the GSO-projection. This further reduces the representation to one copy of $\mathbf{2}$. The only state which has a chance to be a BPS state is the highest weight state of this representation $\mathbf{2}$.

Its mass-shell condition reads as follows:
\be 
-\frac{4j_0(j_0-1)}{k}+\frac{2\ell_0(\ell_0+1)}{2k}+\frac{2k-3}{4k}=\frac{1}{2}\ .
\ee
Here, the first summand comes from the Casimir of $\mathfrak{sl}(2,\mathbb{R})$. The second summand is the standard contribution $h/2$ of the symmetric product orbifold: a state in the seed theory always gives rise to a state in the twisted sector with the same quantum numbers and the conformal weight divided by the twist. The third term is the ground-state energy we explained above. Finally, we equate this with $\tfrac{1}{2}$, which is the normal-ordering constant in the NS sector. Solving the mass-shell condition gives
\be 
j=j_0=\frac{1}{2}\ell_0+\frac{3}{4}=\frac{1}{2}\ell+\frac{1}{2}\ .
\ee
So the state is not BPS.

Let us now consider the R sector. This makes a difference only for the 4 fermions which get orbifolded, which receive a ground-state energy of $-\tfrac{1}{16}$ instead of $\tfrac{1}{16}$. Thus, in comparison to the above groundstate energy, we have to subtract $\tfrac{1}{2}$ and obtain the following ground-state energy of the twisted sector:
\be 
h=\frac{2k-3}{4k}-\frac{1}{2}=-\frac{3}{4k}\ .
\ee
The six unorbifolded fermions are still Ramond-moded. Only four are physical, they are $\psi^\pm$ and $\chi^{+,\pm}+\chi^{-,\pm}$. Hence their zero-modes will generate the representation $(\mathbf{2},\mathbf{2})$ of $\mathfrak{sl}(2,\mathbb{R}) \oplus \mathfrak{su}(2)$. So, the lowest weight state w.r.t.~$\mathfrak{sl}(2,\mathbb{R})$ and the highest weight state w.r.t.~$\mathfrak{su}(2)$ has a chance of being BPS. This state has $j=j_0-\tfrac{1}{2}$ and $\ell=\ell_0+\tfrac{1}{2}$. The mass-shell condition is almost identically as before, it reads:
\be 
-\frac{4j_0(j_0-1)}{k}+\frac{2\ell_0(\ell_0+1)}{2k}-\frac{3}{4k}=0\ .
\ee
Its solution is again $j_0=\tfrac{1}{2}\ell_0+\tfrac{3}{4}$, which this time implies
\be 
j=j_0-\frac{1}{2}=\frac{1}{2}\ell_0+\frac{1}{4}=\frac{1}{2}\ell\ .
\ee

This agrees with the BPS bound \eqref{N3 BPS bound} and so we have found a BPS state in the twisted sector. This state has to be preserved by the GSO-projection, since otherwise the states would not arrange themselves into $\mathcal{N}=3$ multiplets. In fact, we now see that the massless states of the NS sector and the R sector precisely generate the short representation $[j]_\mathrm{S}=[j_0-\tfrac{1}{2}]_\mathrm{S}$ of the global $\mathcal{N}=3$ algebra. Thus, we also manifestly see that all massless states are in fact orbifold-invariant.\footnote{One can again show this directly by using a character argument as we did above for the untwisted sector.}

Now let us now determine the multiplicity of the twisted sector state. It turns out to be 2, as we now explain. The orbifold action has two fixed submanifolds, coming from the fact that the $\mathbb{Z}_2$-action on $\mathrm{S}^1$ has two fixed points. Both of these fixed manifolds are associated with a twisted sector. In fact, the BPS states should precisely correspond to the new cohomology classes of these two singular loci. One can see this also on the character level. For an orbifold of a free compactified boson, the twisted sector has a multiplicity of 2, as discussed e.g.~in \cite{Ginsparg:1988ui}.

Since these are the only states at the massless level, this is the complete list of BPS states. Note that the conformal weight of the BPS states in the twisted sector is in $\tfrac{1}{4}\mathbb{Z}_{> 0}$. There is no dependence on $k$, as it was the case for \cite{Datta:2017ert}. Thus, we expect the theory to be spacetime-supersymmetric for all values of $k$.\footnote{Of course, $k$ is still even due to our choice of conventions.} Hence the BPS spectrum of the twisted sector reads:
\be 
\bigoplus_{\ell \in \frac{1}{2}\mathbb{Z}_{>0}} 2[\ell]_\mathrm{S} \otimes [\ell]_\mathrm{S}\ . \label{worldsheet_twisted_BPS spectrum}
\ee
The lower bound on the spin comes from the relation $\ell=\ell_0+\tfrac{1}{2}$ and $\ell_0 \ge 0$. Note that the $\mathfrak{su}(2)$-spin in the twisted sector can take any half-integer value. This is in contrast to the untwisted sector, where the $\mathfrak{su}(2)$-spin was integer.

\subsection{Summary and comparison}\label{sheet_bps_comp}
We finally conclude that the complete BPS spectrum from string theory in the limit $k \to \infty$ is given by
\be 
\bigoplus_{\ell\in \mathbb{Z}_{\ge 0}}[\ell]_{\mathrm{S}} \otimes [\ell]_{\mathrm{S}}\oplus 2\big( [\ell+\tfrac{1}{2}]_\mathrm{S} \otimes [\ell+\tfrac{1}{2}]_\mathrm{S}\big)\oplus\big([\ell+1]_{\mathrm{S}} \otimes [\ell+1]_{\mathrm{S}}\big) \oplus \bigoplus_{\ell\in \mathbb{Z}_{\ge 0}+\frac{1}{2}} 2\big( [\ell+\tfrac{1}{2}]_\mathrm{S} \otimes [\ell+\tfrac{1}{2}]_\mathrm{S}\big)\ .
\ee
We have written the result in a suggestive way, whose meaning will become apparent below.

In the language of half-integer Hodge-diamonds introduced above, the BPS spectrum reads
\be 
\begin{tabular}{ccc}
&\vdots & \\
0 & & 0 \\
& 2 & \\
0 & & 0 \\
& 2 & \\
0 & & 0 \\
& 1 &
\end{tabular}
\oplus
\begin{tabular}{ccc}
&\vdots & \\
 & 2&  \\
& 2 & \\
 &2 &  \\
& 2 & \\
 &2 &  \\
& 0 &
\end{tabular}=
\begin{tabular}{ccc}
&\vdots & \\
0 & 2& 0 \\
& 4 & \\
0 &2 & 0 \\
& 4 & \\
0 &2 & 0 \\
& 1 &
\end{tabular}\ ,
\ee
where the first and the second terms on the LHS correspond to the $\mathbb Z_2$ untwisted and twisted sectors, respectively. This matches precisely eq. \eqref{SymNS0Z2_bps} and hence gives a strong test of the proposed duality. This also provides another strong test of the recent proposal of \cite{Eberhardt:2017pty} of the large $\mathcal{N}=4$ duality between string theory on $\mathrm{AdS}_3 \times \mathrm{S}^3 \times \mathrm{S}^3  \times \mathrm{S}^1$ and the symmetric orbifold of $\mathcal{S}_\kappa$ in the case of $\kappa=0$. 

When repeating the BPS analysis for finite $k$, we find the same BPS spectrum with upper cutoff in the conformal weight of the dual CFT $\tfrac{c}{12}$, where $c$ is the Brown-Henneaux central charge given in \eqref{Brown Henneaux central charge}. Additionally, there are some BPS states missing. This is much in parallel of the situation for all other $\mathrm{AdS}_3/\mathrm{CFT}_2$ dualities and is discussed in \cite{Seiberg:1999xz, Eberhardt:2017pty}. The upper cutoff matches with what we have concluded at the end of section~\ref{CFT_bps_full}.

\section{Elliptic genus in string theory}\label{sheet_ellgen}

In this section, we will compute the space-time elliptic genus following \cite{Datta:2017ert} and \cite{deBoer:1998us}. There are some additional subtleties arising, as mostly already explained in section \ref{CFT_ellgen}. In spacetime, we compute what we called the ``modified" $\widetilde{{\rm NS}}$ elliptic genus in section~\ref{CFT_ellgen}. We distinguish the $\mathbb{Z}_2$ untwisted and twisted sector contributions.

The elliptic genus is sensitive to all states of the form $|\text{anything} \rangle\otimes |\text{chiral primary}\rangle$. Because of this, we need to know the complete massless spectrum to compute the elliptic genus. From an $\mathcal{N}=3$ perspective, not all massless fields sit in short multiplets, so the comparison is a check of the proposal beyond the BPS spectrum.

\subsection{The $\mathbb{Z}_2$ untwisted sector}\label{sheet_ellgen_Z2U}
We first treat the untwisted sector. We worked out in Section \ref{sheet_bps} the BPS spectrum of the untwisted sector. For convenience, we reproduce here formula \eqref{worldsheet_untwisted_BPS spectrum}:
\be
\bigoplus_{\ell \in \mathbb{Z}_{\ge 0}} \big([\ell]_\mathrm{S} \otimes [\ell]_\mathrm{S}\big) \oplus\big( [\ell+1]_\mathrm{S} \otimes [\ell+1]_\mathrm{S}\big)\ .
\ee
There are further BPS states in the background $\mathrm{AdS}_3 \times \mathrm{S}^3 \times \mathrm{S}^3 \times \mathrm{S}^1$, which are $\mathbb{Z}_2$-odd and are projected out when taking the orbifold. These are the states in \eqref{large N4 BPS spectrum} missing in \eqref{worldsheet_untwisted_BPS spectrum}:
\be 
\bigoplus_{\ell \in \mathbb{Z}_{\ge 0}} \big([\ell+1]_\mathrm{S} \otimes [\ell]_\mathrm{S}\big) \oplus \big([\ell]_\mathrm{S} \otimes [\ell+1]_\mathrm{S}\big)\ .
\ee
In the background $\mathrm{AdS}_3 \times \mathrm{S}^3 \times \mathrm{S}^3 \times \mathrm{S}^1$, these states are highest weight states of short $\mathcal{N}=4$ multiplets.\footnote{Additionally, there are also long $\mathcal{N}=4$ multiplets with $\ell^+ \ne \ell^-$, but these are long for both the left- and right-movers and hence do not contribute to the elliptic genus.} When acting with the $\mathbb{Z}_2$-orbifold, every second state of the short $\mathcal{N}=4$ multiplet is projected out. This is described on the level of the characters in detail in Appendix~\ref{app_alg}. As a consequence, the untwisted string theory elliptic genus can be naturally decomposed in terms of these characters.

The supergravity elliptic genus is defined as
\be 
\mathcal{Z}^{\mathrm{sp}}(z,\tau)=\mathrm{tr}_{\mathrm{NS}\otimes \text{NS chiral primary}}\, \left((-1)^{\mathrm{F}} q^{L_0} y^{J_0} \right)\ . \label{string elliptic genus definition}
\ee
Note that we have not included the ground-state energy $-\tfrac{c}{24}$, since it diverges in the limit $k \to \infty$. This is analogous to what we have done in the CFT computation. $J_0$ denotes the Cartan-generator of the $\mathfrak{su}(2)_k$-algebra. `sp' stands for single-particle, since we restrict to the single-particle sector of supergravity. This is then compared with the single-particle contribution from the CFT.

When restricting to massless fields, we obtain the following contribution from the untwisted sector to the elliptic genus:
\begin{align} 
\mathcal{Z}^{\mathrm{U}_{\mathbb{Z}_2}}(z,\tau)&=\sum_{\ell=0}^\infty \Big(\tilde{\chi}_{\frac{\ell}{2}}^{\mathcal{N}=4,\, (-1)^{\mathrm{F}}}(z,\tau)-\big[\chi_{\frac{\ell}{2},\frac{\ell}{2}}^{\mathcal{N}=4,\, (-1)^{\mathrm{F}}}(z,\tau)-\tilde{\chi}_{\frac{\ell}{2}}^{\mathcal{N}=4,\,(-1)^{\mathrm{F}}}(z,\tau)\big]\nonumber\\
&\qquad+\tilde{\chi}_{\frac{\ell+1}{2}}^{\mathcal{N}=4,\, (-1)^{\mathrm{F}}}(z,\tau)-\big[\chi_{\frac{\ell+1}{2},\frac{\ell+1}{2}}^{\mathcal{N}=4,\, (-1)^{\mathrm{F}}}(z,\tau)-\tilde{\chi}_{\frac{\ell+1}{2}}^{\mathcal{N}=4,\, (-1)^{\mathrm{F}}}(z,\tau)\big]\Big)\ . \label{massless_elliptic genus_untwisted}
\end{align}
The characters $\tilde{\chi}_\ell^{\mathcal{N}=4}(z,\tau)$ are the orbifolded $\mathcal{N}=4$ characters with orbifold even highest weight state as given in Appendix~\ref{app_char_branch}. Since two of the BPS states for each $\ell$ are orbifold even, this contributes the first term in both lines. The terms inside the square brackets come from the orbifold odd highest weight state, in which case the orbifold odd states of the large $\mathcal{N}=4$ multiplet survive. Since we are computing the elliptic genus, we have to insert an additional $(-1)^{\mathrm{F}}$, which we have indicated in the characters. 

When inserting the explicit formulas of the characters as derived in Appendix~\ref{app_alg}, many cancellations occur and we end up with
\be 
\mathcal{Z}^{\mathrm{U}_{\mathbb{Z}_2}}(z,\tau)=\frac{2}{1-yq^{\frac12}}+\frac{2}{1-y^{-1}q^{\frac12}}-3\ ,
\ee
which precisely matches \eqref{ellgen_NSt_oddSym_UZ2}. Note that from a symmetric orbifold perspective, there are no untwisted BPS states in the even twist sectors, so the subtlety with fermion zero modes does not come into play.

\subsection{The $\mathbb{Z}_2$ twisted sector}\label{sheet_ellgen_Z2T}
Recall from \eqref{worldsheet_twisted_BPS spectrum} the BPS spectrum of the twisted sector:
\be 
\bigoplus_{\ell \in \frac{1}{2}\mathbb{Z}_{>0}} 2[\ell]_\mathrm{S} \otimes [\ell]_\mathrm{S}\ . 
\ee
In the twisted sector, the massless states sit only in short $\mathcal{N}=3$ multiplets, as one can see directly from the string theory discussion. Hence the analogue of \eqref{massless_elliptic genus_untwisted} is
\be 
\mathcal{Z}^{\mathrm{T}_{\mathbb{Z}_2}}(z,\tau)=\sum_{\ell \in \frac{1}{2}\mathbb{Z}_{>0}} \big(\epsilon_\ell^{(1)}+\epsilon_\ell^{(2)}\big) \chi_\ell^{\mathcal{N}=3,\,(-1)^{\mathrm{F}}}(z,\tau)\ . \label{twisted elliptic genus i}
\ee
The character formulas for the $\mathcal{N}=3$ algebra can again be found in Appendix~\ref{app_alg} and the $(-1)^{\mathrm{F}}$ is easily inserted.
 
Here, a difficulty arises. We have inserted two constants $\epsilon_\ell^{(1)}$ and $\epsilon_\ell^{(2)}$ for the two BPS states, which take value in $\pm 1$, depending on whether the BPS states are bosonic or fermionic. This is difficult to fix from a string theory perspective and we were not able to do so. Instead we fix the fermion numbers holographically by comparison with the dual CFT. We note that $\ell \in \mathbb{Z}+\tfrac{1}{2}$ corresponds in the dual CFT to odd twist and $\ell \in \mathbb{Z}$ to even twist. We have seen in Section~\ref{CFT_bps} and \ref{CFT_ellgen} that for odd twists the two BPS states are both bosons. On the other hand, for even twist, they have opposite fermion numbers due to the existence of fermion zero-modes and hence cancel out. We thus conclude that the correct choices for the constants $\epsilon_\ell^{(i)}$ are
\be 
\epsilon_\ell^{(1)}=1\ , \quad \epsilon_\ell^{(2)}=(-1)^{2\ell+1}\ .
\ee
Because of this, only terms with $\ell \in \mathbb{Z}+\tfrac{1}{2}$ contribute to the sum \eqref{twisted elliptic genus i}. The $\mathcal{N}=3$ characters can now be inserted and we finally obtain
\begin{align}
\mathcal{Z}^{\mathrm{T}_{\mathbb{Z}_2}}(z,\tau)&=2\sum_{\ell \in \frac{1}{2}\mathbb{Z}_{\ge 0}+\frac{1}{2}} \chi_\ell^{\mathcal{N}=3,\,(-1)^{\mathrm{F}}}(z,\tau) \\
&=\frac{2y^{\frac{1}{2}}q^{\frac{1}{4}}}{1-yq^{\frac{1}{2}}}+\frac{2y^{-\frac{1}{2}}q^{\frac{1}{4}}}{1-y^{-1}q^{\frac{1}{2}}}\ .
\end{align}
This is in perfect agreement with \eqref{ellgen_NSt_oddSym_TZ2}.

\section{Conclusions}\label{conc}
In this paper, we considered string theory on $\mathrm{AdS}_3 \times (\mathrm{S}^3 \times \mathrm{S}^3 \times \mathrm{S}^1)/\mathbb{Z}_2$ and conjectured it to be dual to the symmetric orbifold of the $\mathcal{S}_0/\mathbb{Z}_2$ theory. This proposal is in the spirit of \cite{Datta:2017ert}.

The background we have looked at is interesting for a variety of reasons. In particular, it supports ${\cal N}=(3,3)$ supersymmetry, which allows for a BPS spectrum and a non-vanishing elliptic genus and quantizes the conformal weights of BPS states. Thus the background is sufficiently complicated for indices not to vanish, but easy enough to have good control over protected quantities.

In our proposed duality, we matched the BPS spectrum on both sides and found agreement. This was particularly non-trivial in the twisted sector of the $\mathbb{Z}_2$-orbifold, where new BPS states arise. This matching gives also substantial new evidence for the proposal in \cite{Eberhardt:2017pty} that string theory on $\mathrm{AdS}_3 \times \mathrm{S}^3 \times \mathrm{S}^3 \times \mathrm{S}^1$ should be dual to the symmetric product of $\mathcal{S}_\kappa$ (a generalization of the theory $\mathcal{S}_0$ discussed in this paper), at least in the case when the two spheres have equal size. While in the large ${\cal N}=4$ case the elliptic genus vanishes, it is non-vanishing after taking the orbifold. The comparison allowed us to perform a test of our proposal and of the large ${\cal N}=4$ duality beyond the BPS spectrum.

Let us comment on the r\^ole of higher spin algebras in this duality. In \cite{Eberhardt:2018plx}, the higher spin symmetry in the large ${\cal N}=4$ duality was elucidated. While the symmetric product orbifold of ${\cal S}_0$ does not possess a ${\cal N}=4$ supersymmetric higher spin algebra, one can still define two different kinds of higher spin algebras. First, it supports an ${\cal N}=2$ supersymmetric higher spin algebra, which can be seen by bosonisation of the two uncharged fermions in ${\cal S}_0$ \cite{Eberhardt:2017pty}. This ${\cal N}=2$ higher spin algebra is invariant under the $\mathbb{Z}_2$-orbifold. Another approach was investigated in \cite{Eberhardt:2018plx}, which breaks all supersymmetry, but keeps the R-symmetry explicit. This yields the higher spin algebra $\mathfrak{ho}(4|1)[0]$ in the case of ${\cal S}_0$, which is broken down to $\mathfrak{ho}(3|1)[0]$ when taking the $\mathbb{Z}_2$-orbifold. This shows that the higher spin dualities of \cite{Creutzig:2014ula} are not directly embeddable in the stringy duality, similar to what was concluded in the large ${\cal N}=4$ case for the higher spin duality of \cite{Gaberdiel:2013vva}.

Several directions for future research seem promising. We have mentioned in the main text that two moduli emerge from the twisted sector of the $\mathbb{Z}_2$-orbifold. This indicates that the background geometry might become smooth after turning on these moduli. This would yield an ${\cal N}=3$ supergravity background. To the best of our knowledge, no such background is known in the supergravity literature. 

The new modulus implies also the existence of a non-trivial moduli space of supersymmetric vacua, whose comparison on both sides of the duality would give further evidence for the proposed duality. For string theory on ${\rm AdS}_3 \times {\rm S}^3 \times {\bb T}^4$ and ${\rm AdS}_3 \times {\rm S}^3 \times {\rm K3}$, this was done in \cite{Dijkgraaf:1998gf}.

We have not tried to engineer the background by employing a D-brane construction. This was partly done in \cite{Yamaguchi:1999gb}, but it would be interesting to understand the gauge theory better and show that it has ${\cal N}=(3,3)$ supersymmetry, see \cite{Tong:2014yna}.

One could hope to realise the background as a near-horizon limit of a black hole. Since the modified elliptic genus is non-vanishing, matching black hole entropies would then become possible.

Finally, it would be interesting to study the stringy duality on $\mathrm{AdS}_3\times(\mathrm{S}_3\times\mathrm{S}_3\times\mathrm{S}_1)/\mathbb Z_2$ backgrounds with $\mathcal N=(3,1)$, $(1,3)$, and $(1,1)$ supersymmetries further. The $\mathcal N=1$ theory does not contain a BPS spectrum. However, as discussed in section \ref{CFT_bps_S0Z2}, since the states in Table~\ref{hdgS0Z2untw} are obtained from the action of $\mathbb Z_2$  on chiral primaries of the \l SCA, one still believes that they are protected. It seems natural then to conjecture that the CFTs dual to these configurations are again the symmetric orbifold $\mathrm{Sym}^N(\mathcal S_0/\mathbb Z_2)$ with appropriate supersymmetry content. Furthermore, it would be interesting to determine and compute non-supersymmetric indices which will shed further light on the proposed family of stringy dualities, see e.g.~\cite{Dunne:2018hog} for recent work. We hope to study some of these issues in the near future.

\acknowledgments
We would like to thank Arash Arabi Ardehali, Shouvik Datta, Matthias Gaberdiel, Christoph Keller, Samir Mathur, A.~W.~Peet, Kostas Skenderis, and Marika Taylor for helpful discussions. We thank Matthias Gaberdiel and Christoph Keller for reading the manuscript and Matthias Gaberdiel for comments. LE thanks the Institute for Advanced Study in Princeton for hospitality during the initial stages of this work and Stockholm University for hospitality while the bulk of this work was carried out. IGZ thanks the STAG Research Centre at Southampton University, University of Arizona, Ohio State University, and University of Toronto for hospitality during the final stages of this work and the organisers of the \emph{Workshop on holography, gauge theories and black holes} for the stimulating environment. Our work is supported by the Swiss National Science Foundation through the NCCR SwissMAP.

\appendix
\section{The $\mathcal S_0$ theory}\label{app_S0}
In this appendix we present the free field realisation of the generating currents of the $\mathcal S_0$ theory, see \cite[appendix C]{Eberhardt:2017pty} for the field content of $\mathcal S_\kappa$ theories in general. The $\mathcal S_0$ theory has one real boson $\partial\phi$ and four real fermions $\psi^{\mu\nu}$ and supports the \l SCA. The indices $\mu,\nu\in\{\pm\}$ are bispinor indices of $\mathfrak{su}(2)^+\oplus\mathfrak{su}(2)^-$. The R-symmetry currents $A^{\pm,a}$ are of the form:
\bea
&&A^{+,+}_m=\sum_r\psi^{+-}_r\psi^{++}_{m-r}\ ,\quad A^{+,-}_m=\sum_r\psi^{--}_r\psi^{-+}_{m-r}\ ,\label{S0_A10}\\
&&A^{+,3}_m=\frac12\sum_r(:\psi^{++}_r\psi^{--}_{m-r}:+:\psi^{+-}_r\psi^{-+}_{m-r}:)\ ,\label{S0_A11}\\
&&A^{-,+}_m=\sum_r\psi^{++}_r\psi^{-+}_{m-r}\ ,\quad A^{-,-}_m=\sum_r\psi^{+-}_r\psi^{--}_{m-r}\ ,\label{S0_A12}\\
&&A^{-,3}_m=\frac12\sum_r(:\psi^{++}_r\psi^{--}_{m-r}:-:\psi^{+-}_r\psi^{-+}_{m-r}:)\ .\label{S0_A13}
\eea
where we consider the complex basis $a\in\{\pm,3\}$ which is related to the basis $i\in\{1,2,3\}$ in Appendix \ref{app_alg_lN4} as $A^{\pm,a}\equiv A^{\pm,1}\pm iA^{\pm,2}$. The supercurrents, stress-energy tensor, the $\mathfrak{u}(1)$ current, and the free fermions read
\bea
&G^{++}_r&=\sum_r\psi^{++}_s\alpha_{r-s}+\mathrm{i}\sum_{s,t}:\psi^{++}_t\psi^{+-}_s\psi^{-+}_{r-s-t}:\ ,\label{S0_GTUQ10}\\
&G^{+-}_r&=\sum_r\psi^{+-}_s\alpha_{r-s}-\mathrm{i}\sum_{s,t}:\psi^{++}_t\psi^{+-}_s\psi^{--}_{r-s-t}:\ ,\label{S0_GTUQ11}\\
&G^{-+}_r&=-\sum_r\psi^{-+}_s\alpha_{r-s}+\mathrm{i}\sum_{s,t}:\psi^{++}_t\psi^{--}_s\psi^{-+}_{r-s-t}:\ ,\label{S0_GTUQ12}\\
&G^{--}_r&=\sum_r\psi^{--}_s\alpha_{r-s}+\mathrm{i}\sum_{s,t}:\psi^{+-}_t\psi^{--}_s\psi^{-+}_{r-s-t}:\ ,\label{S0_GTUQ13} \\
&L_n&=\frac12\sum_m:\alpha_m\alpha_{n-m}:+\sum_{r}\left(\tfrac n2-r\right)(:\psi^{++}_r\psi^{--}_{n-r}:+:\psi^{+-}_r\psi^{-+}_{n-r}:)\ ,\label{S0_GTUQ14}\\
&U_m&=\alpha_m\ ,\label{S0_GTUQ15}\\
&Q^{\mu\nu}_r&=\psi^{\mu\nu}_r\ ,\label{S0_GTUQ16}
\eea
where $\alpha_m$ denote the boson modes. The (anti-)commutation relations of these fields reproduce (\ref{A10})-(\ref{A19}) for $k^+=k^-=1$.

\section{Large $\mathcal{N}=4$ and $\mathcal{N}=3$ superalgebras and characters}\label{app_alg}
\subsection{Superconformal algebras}
\subsubsection{Large $\mathcal{N}=4$ SCA}\label{app_alg_lN4}
The large $\mathcal{N}=4$ SCA comes in two guises, commonly referred to as $A_\gamma$ and $\tilde{A}_\gamma$ in the literature. We will use the `linear' version $A_\gamma$. It contains, beyond the four supercurrents, the generators of the R-symmetry $\mathfrak{su}(2)_{k^+} \oplus \mathfrak{su}(2)_{k^-} \oplus \mathfrak{u}(1)$, which we denote by $A^{\pm,i}$ and $U$, respectively, for $i=1,2,3$ denoting adjoint indices of $\mathfrak{su}(2)_{k^\pm}$. In bispinor notation, the algebra reads
\begin{align}
{}[U_m,U_n]  = &  \tfrac{k^+ + k^-}{2} \, m \, \delta_{m,-n}  \label{A10} \\
{}[A^{+, i}_m, Q^{\mu\nu}_r]  = &  \tfrac{1}{2}\tensor{(\sigma^i)}{_\rho^\mu} \, Q^{\rho\nu}_{m+r}   \label{A11} \\
{}[A^{-, i}_m, Q^{\mu\nu}_r]  = &  \tfrac{1}{2}\tensor{(\sigma^i)}{_\rho^\nu} \, Q^{\mu\rho}_{m+r}   \label{A12} \\
{} \{Q^{\mu\nu}_r,Q^{\rho\tau}_s \}  = &  (k^+ + k^-) \, \epsilon^{\mu\rho}\epsilon^{\nu\tau} \, \delta_{r,-s}  \label{A13} \\
{}[A^{\pm, i}_{m}, A^{\pm, j}_{n} ]  = &  \tfrac{k^\pm}{2}\, m \, \delta^{ij}\, \delta_{m,-n} + {\rm i}\, \epsilon^{ijl}\, A^{\pm, l}_{m+n}  \label{A14} \\
{} [U_m,G^{\mu\nu}_r]  = & {\rm i}\,m \, Q^{\mu\nu}_{m+r}  \label{A15} \\
{}[A^{+, i}_{m},G^{\mu\nu}_r]  = &  \tfrac{1}{2}\tensor{(\sigma^i)}{_\rho^\mu} \,G^{\rho\nu}_{m+r}  + (1-\gamma)\, m \, \tensor{(\sigma^i)}{_\rho^\mu}\, Q^{\rho\nu}_{m+r}  \label{A16} \\
{}[A^{-, i}_{m},G^{\mu\nu}_r]  = &  \tfrac{1}{2}\tensor{(\sigma^i)}{_\rho^\nu} \,G^{\mu\rho}_{m+r}  - \gamma\, m \, \tensor{(\sigma^i)}{_\rho^\nu}\, Q^{\mu\rho}_{m+r}  \label{A17} \\
{} \{Q^{\mu\nu}_r,G^{\rho\tau}_s\}  = & 2\,  \epsilon^{\mu\pi}\tensor{(\sigma_i)}{_\pi^\rho}\epsilon^{\nu\tau}\, A^{+, i}_{r+s} - 2 \epsilon^{\nu\pi}\tensor{(\sigma_i)}{_\pi^\tau}\epsilon^{\mu\rho}\, A^{-, i}_{r+s} + 2 {\rm i}\,\epsilon^{\mu\rho}\epsilon^{\nu\tau} \, U_{r+s}  \label{A18} \\
{} \{G^{\mu\nu}_r,G^{\rho\tau}_s\}  = & -\tfrac{2c}{3}\, \epsilon^{\mu\rho}\epsilon^{\nu\tau}\, (r^2 - \tfrac{1}{4}) \delta_{r,-s} - 4\epsilon^{\mu\rho}\epsilon^{\nu\tau}\, L_{r+s} \nonumber \\
 &  + 4\, (r-s)\, \left(\gamma\, \epsilon^{\mu\pi}\tensor{(\sigma^i)}{_\pi^\rho}\epsilon^{\nu\tau}\, A^{+, i}_{r+s} +(1-\gamma)\, \epsilon^{\nu\pi}\tensor{(\sigma^i)}{_\pi^\tau}\epsilon^{\mu\rho} A^{-, i}_{r+s} \right) \ . \label{A19}
\end{align}
The central charge $c$ and the parameter $\gamma$ are given by
\begin{equation}\label{A9}
\gamma = \frac{k^-}{k^+ + k^-} \ , \qquad c = \frac{6 k^+ k^-}{k^+ + k^-} \ .
\end{equation}
Here, greek indices $\mu,\nu,\dots$ are spinor indices and get as usual raised and lowered by the epsilon symbol $\epsilon_{\mu\nu}$, which we have indicated explicitly. $\sigma^i$ denotes the Pauli matrices, i.e.~the two-dimensional spinor representation of $\mathfrak{su}(2)$.

\subsubsection{$\mathcal{N}=3$ SCA}\label{app_alg_N3}
The $\mathcal{N}=3$ SCA also has a linear and a non-linear version \cite{Goddard:1988wv}. In analogy with the case of the \l SCA discussed in \ref{app_alg_lN4}, we are interested in the linear version of the algebra (see appendix \ref{app_reps_N3} for more details). The $\mathcal{N}=3$ SCA is a subalgebra of $A_\gamma$ for $\gamma=\tfrac{1}{2}$. Using equation (\ref{A9}), this yields $k^+=k^-$. We define $k \equiv k^++k^-=2k^+=2k^-$.

The generators of the linear $\mathcal{N}=3$ algebra are $\{G^i, A^i,Q\}$. The three supercurrents, $G^i$, are related to the supercurrents of the \l SCA as
\be\label{N3lN4Gs}
G^i_r \equiv \tfrac{1}{2}(\sigma^i)_{\mu\nu}G_r^{\mu\nu}.
\ee
As for the R-symmetry group, we keep the diagonal $\mathfrak{su}(2)_k$ whose associated currents are
\be\label{N3lN4As}
A^i \equiv A^{+,i}+A^{-,i}.
\ee
The index $i\in\{1,2,3\}$ corresponds to the adjoint representation of $\mathfrak{su}(2)_k$. Finally, we keep the free fermion
\be\label{N3lN4Q}
Q \equiv \frac{1}{2}\epsilon_{\mu\nu} Q^{\mu\nu}_r=\tfrac{1}{2}(Q^{+-}_r-Q^{-+}_r).
\ee
The commutation relations read 
\begin{align}
{}[A^{i}_m, Q_r]  = &  0   \label{A20} \\
{} \{Q_r,Q_s \}  = &  \tfrac{k}{2}\, \delta_{r,-s}  \label{A21} \\
{}[A^{i}_{m}, A^{j}_{n} ]  = &  \tfrac{k}{2}\, m \, \delta^{ij}\, \delta_{m,-n} + {\rm i}\, \epsilon^{ijl}\, A^{\pm, l}_{m+n}  \label{A22} \\
{}[A^{i}_{m},G^{j}_r]  = &  \mathrm{i}\epsilon^{ijk} G^k_{m+r}+m\delta^{ij} Q_{m+r}  \label{A23} \\
{} \{Q_r,G^{i}_s\}  = & A^i_{r+s}\ , \label{A24} \\
{} \{G^{i}_r,G^{j}_s\}  = & \tfrac{k}{2}\, (r^2 - \tfrac{1}{4}) \delta^{ij}\delta_{r,-s} 
+2\delta^{ij}\, L_{r+s}   +  (r-s)\, \mathrm{i}\epsilon^{ijk}\, A^{k}_{r+s}  \ .\label{A25}
\end{align}
The central charge is then $c=\tfrac{3}{2}k$.

\subsection{Representations}\label{app_reps}
\subsubsection{Large $\mathcal{N}=4$ SCA}\label{app_reps_lN4}
The representations of the large $\mathcal{N}=4$ SCA are labelled by the quantum numbers $(h,\ell^+,\ell^-,u)$, which correspond to the conformal weight, the two $\mathfrak{su}(2)$ spins, and the $\mathfrak{u}(1)$ charge, respectively. In terms of these, the BPS bound reads
\be 
h \ge \frac{k^-\ell^++k^+\ell^-+(\ell^+-\ell^-)^2+u^2}{k^++k^-}\ . \label{N4 BPS bound}
\ee
For more details, see e.g.~\cite{Gaberdiel:2013vva}.

\subsubsection{$\mathcal{N}=3$ SCA}\label{app_reps_N3}
The BPS bound for the linear $\mathcal N=3$ SCA is the same as that of the non-linear ${\cal N}=3$ algebra and reads
\be 
h \ge \frac{\ell}{2}\ . \label{N3 BPS bound}
\ee 
The representations of the $\mathcal{N}=3$ SCA are labelled by $(h,\ell)$. Notice that this BPS bound means in particular that the conformal weight of BPS states takes quarter-integer values.

When realizing the $\mathcal{N}=3$ SCA as a subalgebra of the large $\mathcal{N}=4$ SCA with $k^+=k^-$, we can decompose a representation $(h,\ell^+,\ell^-,u)$ into $\mathcal{N}=3$ representations. In particular, we obtain the representation $(h,\ell^++\ell^-)$ on the ground state. Note that the $\mathcal{N}=3$ BPS bound does not agree with the $\mathcal{N}=4$ BPS bound except for $\ell^+=\ell^-$ and $u=0$.

\subsection{The global subalgebras}\label{app_alg_wedge}
We will now discuss the global (or `wedge') subalgebras of the relevant superconformal algebras. These are generated by all modes annihilating the in-, as well as the out-vacuum. These are the modes with mode numbers $-h<m<h$, where $h$ is the conformal weight of the respective field. In particular, the free fermions are invisible in the global subalgebra. The $\mathfrak{u}(1)$-current $U_0$ of the large $\mathcal{N}=4$ SCA becomes central and decouples from the algebra. The resulting global subalgebra is known as the exceptional Lie superalgebra $\mathfrak{d}(2,1;\alpha)$, where the parameter $\alpha$ is related to $\gamma$ by
\be 
\alpha=\frac{\gamma}{1-\gamma}\ .
\ee
In the important case $\gamma=\tfrac{1}{2}$ or $\alpha=1$, we have another description thanks to the isomorphism $\mathfrak{d}(2,1;\alpha=1)\cong \mathfrak{osp}(4|2)$. The field content of this Lie superalgebra is seen to be correct, the bosonic subalgebra is $\mathfrak{sp}(2)\oplus\mathfrak{so}(4)\cong \mathfrak{su}(2) \oplus \mathfrak{su}(2) \oplus \mathfrak{su}(2)$. The first $\mathfrak{su}(2)$ is generated by the energy-momentum tensor (or rather its non-compact version $\mathfrak{sl}(2,\mathbb{R})$). The other two $\mathfrak{su}(2)$'s describe the R-symmetry. The fermions transform in the representation $(\mathbf{2},\mathbf{4}) \cong (\mathbf{2},\mathbf{2},\mathbf{2})$ of these algebras. This is indeed the correct transformation behaviour of the supercharges.

The $\mathbb{Z}_2$-quotient we performed above can also be seen on the global superalgebra. It corresponds to interchanging the two $\mathfrak{su}(2)$-factors of $\mathfrak{so}(4)$. The fixed point algebra is then given by $\mathfrak{osp}(3|2)$, which is the global subalgebra of the $\mathcal{N}=3$ SCA. The bosonic and fermionic field content matches again the one we listed above. 

As an aside, we mention the following two facts. We may ask whether there are other values of $\alpha$ for which $\mathfrak{d}(2,1;\alpha)$ has an outer automorphism by which we can orbifold. These outer automorphisms of $\mathfrak{d}(2,1;\alpha)$ are listed in \cite{Frappat:1996pb}. When requiring in addition that $\alpha\in \mathbb{Q}_{\ge 0}$, in which case we can extend the algebra to the large $\mathcal{N}=4$ SCA a unitary manner, only $\alpha=1$ has the non-trivial outer automorphism group $\mathbb{Z}_2$ by which we orbifolded above.

The construction generalizes however to higher rank. The Lie superalgebra $\mathfrak{osp}(2m|2n)$ has still outer automorphism group $\mathbb{Z}_2$. This outer automorphism can be described by the Adjoint representation of an element of superdeterminant $-1$ in $\mathrm{OSP}(2m|2n)$ on its Lie algebra. If we orbifold by this $\mathbb{Z}_2$, the fixed point algebra becomes $\mathfrak{osp}(2m-1|2n)$ \cite{Frappat:1987ix}. For $n=1$, this was discussed in \cite{Creutzig:2014ula}. The case $n>1$ might be relevant for higher spin algebras with extended supersymmetry.

\subsection{$\mathfrak{d}(2,1;\alpha)$ characters}\label{app_char_lN4}
In this subsection, we will discuss the characters of the global large $\mathcal{N}=4$ superalgebra $\mathfrak{d}(2,1;\alpha)$. In our situation of interest, we only need $\alpha=1$, but the characters are generically independent of the value of $\alpha$.\footnote{This is of course not the case for the superconformal characters described in detail in \cite{Petersen:1989zz,Petersen:1989pp}.} We have two chemical potentials $z_\pm$ (with $y_\pm =\mathrm{e}^{2\pi\mathrm{i} z_\pm}$ associated to the two $\mathfrak{su}(2)^\pm$'s). Furthermore, we have the usual chemical potential $\tau$ (with $q=\mathrm{e}^{2\pi\mathrm{i}\tau}$) associated to the third $\mathfrak{su}(2)$, which is the M\"obius subalgebra of the Virasoro algebra. We shall denote an $\mathfrak{su}(2)$-character by $\chi_\ell(z)$:
\be 
\chi_\ell(z)=\frac{y^{\ell+\frac{1}{2}}-y^{-\ell-\frac{1}{2}}}{y^{\frac{1}{2}}-y^{-\frac{1}{2}}}\ .
\ee
A long $\mathfrak{d}(2,1;\alpha)$-multiplet has the form shown in Table~\ref{tab:long N4 multiplet}.
\begin{table}
\begin{center} 
\setlength{\tabcolsep}{-15pt}
\bgroup
\def\arraystretch{1.5}
\begin{tabular}{l|ccccccccccccccc}
\hspace{.5cm}$h+2$ \hspace{.5cm} & & & & & $(\ell^+,\ell^-)$ \\
\hspace{.5cm}$h+\tfrac{3}{2}$& & $(\ell^++\tfrac{1}{2},\ell^-+\tfrac{1}{2})$ & & $(\ell^++\tfrac{1}{2},\ell^--\tfrac{1}{2})$ & & $(\ell^+-\tfrac{1}{2},\ell^-+\tfrac{1}{2})$ & & $(\ell^+-\tfrac{1}{2},\ell^--\tfrac{1}{2})$ \\
\hspace{.5cm}$h+1$ & \hspace{.5cm} $(\ell^++1,\ell^-)$ & & \hspace{.1cm}$(\ell^+,\ell^-+1)$ & & \hspace{.3cm}$2(\ell^+,\ell^-)$\hspace{.3cm} & & $(\ell^+-1,\ell^-)$\hspace{.1cm} & & $(\ell^+,\ell^--1)$\\
\hspace{.5cm}$h+\tfrac{1}{2}$ \hspace{.5cm}& & $(\ell^++\tfrac{1}{2},\ell^-+\tfrac{1}{2})$ & & $(\ell^++\tfrac{1}{2},\ell^--\tfrac{1}{2})$ & & $(\ell^+-\tfrac{1}{2},\ell^-+\tfrac{1}{2})$ & & $(\ell^+-\tfrac{1}{2},\ell^--\tfrac{1}{2})$ \\
\hspace{.5cm}$h$& & & & & $(\ell^+,\ell^-)$
\end{tabular}
\egroup
\setlength{\tabcolsep}{6pt}
\caption{The multiplet structure of a long $\mathfrak{d}(2,1;\alpha)$-multiplet.} \label{tab:long N4 multiplet}
\end{center}
\end{table}
Thus, its character reads:
\begin{align}
\chi_{h,\ell^+,\ell^-}^{\mathcal{N}=4}(z_\pm,\tau)&=\frac{q^h}{1-q} \Big( \big(1+q^2\big) \chi_{\ell^+}(z_+)\chi_{\ell^-}(z_-)+\big(q^{\frac{1}{2}}+q^{\frac{3}{2}}\big)\big(\chi_{\ell^++\frac{1}{2}}(z_+)\chi_{\ell^-+\frac{1}{2}}(z_-)\nonumber\\
&\qquad+\chi_{\ell^++\frac{1}{2}}(z_+)\chi_{\ell^--\frac{1}{2}}(z_-)+\chi_{\ell^+-\frac{1}{2}}(z_+)\chi_{\ell^-+\frac{1}{2}}(z_-)\nonumber\\
&\qquad+\chi_{\ell^+-\frac{1}{2}}(z_+)\chi_{\ell^--\frac{1}{2}}(z_-)\big)\nonumber\\
&\qquad+q\big(\chi_{\ell^++1}(z_+)\chi_{\ell^-}(z_-)+\chi_{\ell^+}(z_+)\chi_{\ell^-+1}(z_-)+\chi_{\ell^+-1}(z_+)\chi_{\ell^-}(z_-)\nonumber\\
&\qquad+\chi_{\ell^+}(z_+)\chi_{\ell^--1}(z_-)+2\chi_{\ell^+}(z_+)\chi_{\ell^-}(z_-)\big)\Big) \\
&=\frac{q^h \chi_{\ell^+}(z_+)\chi_{\ell^-}(z_-)}{1-q}\prod_{\epsilon_+,\epsilon_-=\pm}\big(1+y_+^{\frac{1}{2}\epsilon_+} y_-^{\frac{1}{2}\epsilon_-} q^{\frac{1}{2}}\big)\ .\label{lN4_long_ii}
\end{align}
In the last formulation, the action of the four supercharges is made manifest. There are some exceptions to the multiplet structure displayed in Table \ref{tab:long N4 multiplet} when either $\ell^+ < 1$ or $\ell^- < 1$, according to the tensor product rules of $\mathfrak{su}(2)$-representations. However, eq.~(\ref{lN4_long_ii}) remains true even for low spins.

Similarly, a short multiplet has the structure shown in Table~\ref{tab:short N4 multiplet} with a character
\begin{table}
\begin{center}
\bgroup
\def\arraystretch{1.5}
\begin{tabular}{l|ccccccccccccccc}
$h+\tfrac{3}{2}$& & $(\ell^+-\tfrac{1}{2},\ell^--\tfrac{1}{2})$ \\
$h+1$ &   $(\ell^+,\ell^-)$\hspace{.2cm} & $(\ell^+-1,\ell^-)$ & $(\ell^+,\ell^--1)$\\
$h+\tfrac{1}{2}$ & $(\ell^++\tfrac{1}{2},\ell^--\tfrac{1}{2})$ & $(\ell^+-\tfrac{1}{2},\ell^-+\tfrac{1}{2})$ & $(\ell^+-\tfrac{1}{2},\ell^--\tfrac{1}{2})$ \\
$h$& & $(\ell^+,\ell^-)$
\end{tabular}
\egroup
\caption{The multiplet structure of a short $\mathfrak{d}(2,1;\alpha)$-multiplet.} \label{tab:short N4 multiplet}
\end{center}
\end{table}
\begin{align} 
\chi^{\mathcal{N}=4}_{\ell^+,\ell^-}(z_\pm,\tau)&=\frac{q^{h_{\mathrm{BPS}}}}{1-q}\Big(\chi_{\ell^+}(z_+)\chi_{\ell^-}(z_-)+q^{\frac{1}{2}}\big(\chi_{\ell^+-\frac{1}{2}}(z_+)\chi_{\ell^--\frac{1}{2}}(z_-)\nonumber\\
&\qquad\qquad+\chi_{\ell^+-\frac{1}{2}}(z_+)\chi_{\ell^-+\frac{1}{2}}(z_-)+\chi_{\ell^++\frac{1}{2}}(z_+)\chi_{\ell^--\frac{1}{2}}(z_-)\big)\nonumber\\
&\qquad+q\big(\chi_{\ell^+-1}(z_+)\chi_{\ell^-}(z_-)+\chi_{\ell^+}(z_+)\chi_{\ell^--1}(z_-)\nonumber\\
&\qquad\qquad+\chi_{\ell^+}(z_+)\chi_{\ell^-}(z_-)\big)+q^{\frac{3}{2}} \chi_{\ell^+-\frac{1}{2}}(z_+)\chi_{\ell^--\frac{1}{2}}(z_-)\Big)\\
&=q^{h_{\mathrm{BPS}}}\!\!\!\!\!\!\sum_{\eta_+,\eta_-=\pm 1} \frac{\eta_+ \eta_-y_+^{\eta_+(\ell^++\frac{1}{2})}y_-^{\eta_-(\ell^-+\frac{1}{2})}}{(1-q)(y_+^{\frac{1}{2}}-y_+^{\frac{1}{2}})(y_-^{\frac{1}{2}}-y_-^{\frac{1}{2}})}\!\!\!\prod_{\epsilon_+,\epsilon_-=\pm \atop (\epsilon_+,\epsilon_-) \ne (\eta_+,\eta_-)}\!\!\!\!\!\!\!\big(1+y_+^{\frac{1}{2}\epsilon_+} y_-^{\frac{1}{2}\epsilon_-} q^{\frac{1}{2}}\big)\ . \label{short N4 character}
\end{align}
The conformal weight saturates the BPS bound 
\be 
h_{\mathrm{BPS}}(\ell^+,\ell^-)=\frac{k^-\ell^++k^+\ell^-}{k^++k^-}\ .
\ee
This is the $k^\pm \to \infty$ limit of \eqref{N4 BPS bound}, as befits the BPS bound of the global subalgebra. In the formulation \eqref{short N4 character}, one can see that one of the supercharges always acts trivially. This formula also holds true for small spin $\ell^+,\ell^-$. The multiplet structure of these exceptional cases is treated in \cite{deBoer:1999gea}.

\subsection{$\mathcal{N}=3$ characters}\label{app_char_N3}
We repeat a similar analysis for the characters of $\mathfrak{osp}(3|2)$, the global algebra of the $\mathcal{N}=3$ SCA. A long multiplet has the structure displayed in table~\ref{tab:long N3 multiplet}. Now states arrange themselves into $\mathfrak{su}(2)$-multiplets. We denote the chemical potential of the $\mathfrak{su}(2)$ by $z$.
\begin{table}
\begin{center}
\bgroup
\def\arraystretch{1.5}
\begin{tabular}{l|ccccccccccccccc}
$h+\tfrac{3}{2}$& & $\ell$ \\
$h+1$ & $\ell-1$ & $\ell$ & $\ell+1$ \\
$h+\tfrac{1}{2}$ & $\ell-1$ & $\ell$ & $\ell+1$ \\
$h$& & $\ell$
\end{tabular}
\egroup
\caption{The multiplet structure of a long $\mathfrak{osp}(3|2)$-multiplet.} \label{tab:long N3 multiplet}
\end{center}
\end{table}
The the long character $\mathcal{N}=3$ hence equals
\begin{align}
\chi^{\mathcal{N}=3}_{h,\ell}(z,\tau)&=\frac{q^h}{1-q} \Big(\big(\chi_\ell(z)+q^{\frac{3}{2}}\big)+\big(q^{\frac{1}{2}}+q\big)\big(\chi_{\ell+1}(z)+\chi_\ell(z)+\chi_{\ell-1}(z)\big)\Big) \\
&=\frac{q^h\chi_\ell(z)}{1-q}\big(1+q^{\frac{1}{2}}\big) \prod_{\epsilon= \pm } \big(1+y^\epsilon q^{\frac{1}{2}} \big)\ .
\end{align}
The factorized form again encompasses also low-lying special cases. 

Finally, a short representation of $\mathfrak{osp}(3|2)$ has the form displayed in table~\ref{tab:short N3 multiplet}. 
\begin{table}
\begin{center}
\bgroup
\def\arraystretch{1.5}
\begin{tabular}{l|ccccccccccccccc}
$h+1$ & & $\ell-1$ & &  \\
$h+\tfrac{1}{2}$ & $\ell-1$ & & $\ell$  \\
$h$& &  $\ell$
\end{tabular}
\egroup
\caption{The multiplet structure of a short $\mathfrak{osp}(3|2)$-multiplet.} \label{tab:short N3 multiplet}
\end{center}
\end{table}
Hence, its character reads
\begin{align}
\chi^{\mathcal{N}=3}_{\ell}(z,\tau)&=\frac{q^{h_{\mathrm{BPS}}}}{1-q} \Big(\chi_\ell(z)+q^{\frac{1}{2}}\big(\chi_\ell(z)+\chi_{\ell-1}(z)\big)+q \chi_{\ell-1}(z)\Big) \\
&=q^{h_{\mathrm{BPS}}}\sum_{\eta=\pm}\frac{\eta y^{\eta(\ell+\frac{1}{2})}}{(1-q)(y^{\frac{1}{2}}-y^{-\frac{1}{2}})}\big(1+q^{\frac{1}{2}}\big) \prod_{\epsilon= \pm \atop \epsilon \ne \eta} \big(1+y^\epsilon q^{\frac{1}{2}} \big)\ .
\end{align}
Again, one of the supercharges acts trivially. The BPS bound $h_{\mathrm{BPS}}(\ell)=\tfrac{\ell}{2}$ is saturated. This again encompasses low-lying special cases.

\subsection{The branching $\mathfrak{osp}(4|2) \longrightarrow \mathfrak{osp}(3|2)$}\label{app_char_branch}
As explained in the previous subsection, the global $\mathcal{N}=3$ algebra $\mathfrak{osp}(3|2)$ is a natural subalgebra of the global large $\mathcal{N}=4$ algebra $\mathfrak{d}(2,1;\alpha=1) \cong \mathfrak{osp}(4|2)$. Hence, there should be an associated branching rule, which we can determine on the level of the characters. Since, the subalgebra $\mathfrak{osp}(3|2)$ preserves the diagonal $\mathfrak{su}(2)$, we set the chemical potentials $z \equiv z_+=z_-$. From the expressions of the characters we gave above, the branching rules for a long $\mathfrak{osp}(4|2)$ representation are
\be 
\chi_{h,\ell^+,\ell^-}^{\mathcal{N}=4}(z,\tau)=\sum_{\ell=|\ell^+-\ell^-|,\, \ell^++\ell^--\ell \in \mathbb{Z}}^{\ell^++\ell^-} \big(\chi_{h,\ell}^{\mathcal{N}=3}(z,\tau)+\chi_{h+\frac{1}{2},\ell}^{\mathcal{N}=3}(z,\tau)\big)\ . \label{N4N3 branching long}
\ee
These branching rules are more or less directly inherited from the $\mathfrak{su}(2)$ tensor product decomposition. For a short $\mathfrak{osp}(4|2)$-multiplet, we have
\be 
\chi_{\ell^+,\ell^-}^{\mathcal{N}=4}(z,\tau)=\sum_{\ell=|\ell^+-\ell^-|,\ell^++\ell^--\ell \in \mathbb{Z}}^{\ell^++\ell^-} \chi_{h_{\mathrm{BPS}}(\ell^+,\ell^-),\ell}^{\mathcal{N}=3}(z,\tau)\ . \label{N4N3 branching short}
\ee
The top summand is a short multiplet. Indeed, it has $h=\tfrac{1}{2}\ell^++\tfrac{1}{2}\ell^-=\tfrac{1}{2}\ell$ and hence saturates the $\mathcal{N}=3$ BPS bound.

\subsection{The orbifold action on the characters}\label{app_char_Z2}
We have seen above that the $\mathfrak{osp}(3|2)$-subalgebra may be described by the fixed point set of an involution acting on $\mathfrak{osp}(4|2)$, which interchanges the two $\mathfrak{su}(2)$-factors of $\mathfrak{so}(4) \cong \mathfrak{su}(2) \oplus \mathfrak{su}(2)$. Correspondingly, there is a $\mathbb{Z}_2$-action on the representations of $\mathfrak{osp}(4|2)$ and so on the characters. It is clear that a character $\chi^{\mathcal{N}=4}_{h,\ell^+,\ell^-}$ is mapped to $\chi^{\mathcal{N}=4}_{h,\ell^-,\ell^+}$ under the $\mathbb{Z}_2$-action. Hence, for $\ell^+\ne \ell^-$, the fixed points are simply symmetric or antisymmetric combinations of the two representations. The question becomes much more interesting for $\ell^+=\ell^-$, which is mapped to itself under the orbifold action. $\mathfrak{su}(2) \oplus \mathfrak{su}(2)$-representations $(\ell^+,\ell^-)$ are again mapped to $(\ell^-,\ell^+)$ under the orbifold action. This is the reason why they always have to appear in pairs in the $\mathcal{N}=4$ characters. For the $\mathfrak{su}(2) \oplus \mathfrak{su}(2)$-representation $(\ell,\ell)$, the orbifold even part is seen to be
\be 
\frac{1}{2}\chi_\ell(z)\chi_\ell(z)+\frac{1}{2}\chi_\ell(2z)\ ,
\ee
whereas the orbifold odd part is
\be 
\frac{1}{2}\chi_\ell(z)\chi_\ell(z)-\frac{1}{2}\chi_\ell(2z)\ .
\ee

We have set the chemical potentials equal, since the orbifold-fixed part transforms only under the diagonal $\mathfrak{su}(2)$. Remember also that one uncharged supercharge is orbifold odd, whereas the other three supercharges are orbifold even. Thus, the invariant part of a long $\mathfrak{osp}(4|2)$-character $\chi^{\mathcal{N}=4}_{h,\ell,\ell}$ is
\begin{align}
\widetilde{\chi}^{\mathcal{N}=4}_{h,\ell}(z,\tau)&=\frac{q^h}{1-q} \Big(\tfrac{1}{2}\big(1+q^2\big)\chi_{\ell}(z)\chi_{\ell}(z)+\tfrac{1}{2} \big(1-q^2\big)\chi_\ell(2z)\nonumber\\
&\qquad+\big(q^{\frac{1}{2}}+q^{\frac{3}{2}}\big)\big(\tfrac{1}{2}\chi_{\ell+\frac{1}{2}}(z)\chi_{\ell+\frac{1}{2}}(z)+\chi_{\ell+\frac{1}{2}}(z)\chi_{\ell^-\frac{1}{2}}(z)+\tfrac{1}{2}\chi_{\ell-\frac{1}{2}}(z)\chi_{\ell-\frac{1}{2}}(z)\big)\nonumber\\
&\qquad+\tfrac{1}{2}\big(q^{\frac{1}{2}}-q^{\frac{3}{2}}\big)(\chi_{\ell-\frac{1}{2}}(2z)+\chi_{\ell+\frac{1}{2}}(2z)\big)\nonumber\\
&\qquad+q\big(\chi_{\ell+1}(z)\chi_{\ell}(z)+\chi_{\ell-1}(z)\chi_{\ell}(z)+\chi_{\ell}(z)\chi_{\ell}(z)\big)\Big) \\
&=q^h\frac{\big(1+q^{\frac{1}{2}}\big)\chi_{\ell}(z)\chi_{\ell}(z)+\big(1-q^{\frac{1}{2}}\big)\chi_{\ell}(2z)}{2(1-q)}\prod_{\epsilon=0,\pm}\big(1+y^\epsilon q^{\frac{1}{2}}\big)\ .
\end{align}
In the last formulation, we see again the three orbifold even supercharges manifestly. The last supercharge is orbifold odd, which is why we have to keep also the orbifold odd part of the $\mathfrak{su}(2)\oplus \mathfrak{su}(2)$-multiplet, when it is applied. 

The corresponding formula for a short $\mathcal{N}=4$ representation reads
\begin{align}
\widetilde{\chi}^{\mathcal{N}=4}_{\ell}(z,\tau)&=\frac{q^h}{1-q} \Big(\tfrac{1}{2}\chi_{\ell}(z)\chi_{\ell}(z)+\tfrac{1}{2} \chi_\ell(2z)+q^{\frac{1}{2}}\big(\chi_{\ell+\frac{1}{2}}(z)\chi_{\ell-\frac{1}{2}}(z)+\tfrac{1}{2}\chi_{\ell-\frac{1}{2}}(z)\chi_{\ell-\frac{1}{2}}(z)\nonumber\\
&\qquad+\tfrac{1}{2}\chi_{\ell-\frac{1}{2}}(2z)\big)+q\big(\chi_{\ell-1}(z)\chi_{\ell}(z)+\tfrac{1}{2}\chi_{\ell}(z)\chi_{\ell}(z)-\tfrac{1}{2}\chi_{\ell}(2z)\big)\nonumber\\
&\qquad+\tfrac{1}{2}q^{\frac{3}{2}}(\chi_{\ell-\frac{1}{2}}(z)\chi_{\ell-\frac{1}{2}}(z)-\chi_{\ell-\frac{1}{2}}(2z))\Big) \\
&=\frac{q^h\big(1+q^{\frac{1}{2}}\big)}{2(1-q)\big(y^{\frac{1}{2}}-y^{-\frac{1}{2}}\big)}\Bigg(\sum_{\eta=\pm}\frac{y^{\eta(2\ell+\frac{3}{2})}\big(1+y^{-\eta} q^{\frac{1}{2}}\big)^2}{y-y^{-1}}-\frac{\prod_{\epsilon=\pm}\big(1+y^\epsilon q^{\frac{1}{2}}\big)}{y^{\frac{1}{2}}-y^{-\frac{1}{2}}}\Bigg)\ .
\end{align}
Here, the interpretation of this formula becomes less clear. It holds however again also for low spins.

Finally, we check that this procedure of taking the orbifold even part preserves the $\mathcal{N}=3$ decomposition \eqref{N4N3 branching long} and \eqref{N4N3 branching short}. We have the following formulas for the long and short characters, respectively:
\begin{align}
\widetilde{\chi}^{\mathcal{N}=4}_{h,\ell}(z,\tau)&=\sum_{\ell'=0,\, 2\ell-\ell' \in 2\mathbb{Z}}^{2\ell} \chi_{h,\ell}^{\mathcal{N}=3}(z,\tau)+\sum_{\ell'=0,\, 2\ell-\ell' \in 2\mathbb{Z}+1}^{2\ell}\chi_{h+\frac{1}{2},\ell}^{\mathcal{N}=3}(z,\tau)\ , \label{N4N3 orbifold branching long}\\
\widetilde{\chi}^{\mathcal{N}=4}_{\ell}(z,\tau)&=\sum_{\ell'=0,\, 2\ell-\ell' \in 2\mathbb{Z}}^{2\ell} \chi_{h_{\mathrm{BPS}},\ell}^{\mathcal{N}=3}(z,\tau)\ . \label{N4N3 orbifold branching short}
\end{align}
In the second line, the top component of the summand is again a short character.

\section{Symmetric orbifold of $\mathcal S_0/\mathbb Z_2$}\label{app_SymN}
\subsection{BPS spectrum of $\mathrm{Sym}^N(\mathcal S_0/\mathbb Z_2)$}\label{app_SymN_bps}
\subsubsection{Odd twisted sector of $\mathrm{Sym}^N(\mathcal S_0/\mathbb Z_2)$}\label{app_SymN_bps_odd}
\textbf{${\bf{\mathbb Z_2}}$ untwisted sector:}

The ground state has conformal dimension and R-charge \cite{Lunin:2001pw}:
\be\label{SymNoddgs}
h=\bar h=\frac{c}{24}\Big(n-\frac1n\Big)=\frac{1}{8}\Big(n-\frac1n\Big)\ ,\qquad\qquad\ell=\bar\ell=0\ .
\ee
To construct BPS states, we act on the ground state with fermionic excitations to increase the conformal dimension and the $\mathfrak u(1)$-charge such that they saturate the BPS bound $h=\tfrac\ell2$. The only field which brings us closer to the BPS bound is $\psi^{++}_r$. Since fermions are fractionally moded in the symmetric orbifold twisted sector, we can construct the following state:
\be\label{SymNS0Z2oddn_i}
\psi_{-\frac{1}{2}+\frac{1}{n}}^{++}\;\psi_{-\frac{1}{2}+\frac{2}{n}}^{++}\cdots\psi_{-\frac{1}{2}+\frac{n-1}{2n}}^{++}\;
\tilde{\psi}_{-\frac{1}{2}+\frac{1}{n}}^{++}\;\tilde{\psi}_{-\frac{1}{2}+\frac{2}{n}}^{++}\cdots\tilde{\psi}_{-\frac{1}{2}+\frac{n-1}{2n}}^{++}|\sigma\rangle\ ,
\ee
where $|\sigma\rangle$ is the symmetric orbifold ground state in the twist-$n$ sector and $\tilde\psi$ correspond to the right-moving fermions. This state has conformal weight and R-charge
\be\label{SymNS0Z2oddn_ii}
h=\bar{h}=\frac{1}{8}\Big(n-\frac1n\Big)+\sum_{k=1}^{\frac{n-1}{2}} \left(\frac{1}{2}-\frac{k}{n} \right)=\frac{n-1}{4}\ ,\qquad\qquad\ell=\bar\ell=\frac{n-1}2\ ,
\ee
and so is a BPS state. Moreover, it is orbifold even since we have an even number of fermionic excitations ($n\in\mathbb Z_{\mathrm{odd}>0}$). On top of this BPS state, we can apply $\psi^{++}_{-1/2} \bar{\psi}^{++}_{-1/2}$ to obtain another BPS state. These contribute to BPS states with $h=\bar h=\tfrac\ell2=\tfrac{\bar\ell}2=\tfrac{n+1}4$. We cannot apply only one fermion since the resulting state would not be orbifold even.

\vspace{10pt}\hspace{-25pt}
\textbf{${\bf{\mathbb Z_2}}$ twisted sector:}

In the $\mathbb{Z}_2$-twisted sector, each boson and each fermion contribute to a factor of $q^{\frac1{16}}$ to the ground-state energy in the NS sector. The seed theory has then the ground-state energy $\tfrac{1}{16}+3 \times \tfrac{1}{16}=\tfrac{1}{4}$ (the factor of 3 corresponds to the fact that 3 out of 4 free fermions are orbifolded under the $\mathbb Z_2$). Thus, in the $n^{\mathrm{th}}$ twisted sector of the symmetric orbifold, we have an additional contribution of $\tfrac{1}{4n}$ to the ground-state energy. The zero modes shift the $\mathfrak{u}(1)$-charge by $\tfrac{1}{2}$. On top of this, we apply now the following fermionic oscillators:
\be\label{SymNS0Z2oddn_iii}
\psi_{-\frac{1}{n}}^{++}\;\psi_{-\frac{2}{n}}^{++}\cdots\psi_{-\frac{n-1}{2n}}^{++}\;\tilde{\psi}_{-\frac{1}{n}}^{++}\;\tilde{\psi}_{-\frac{2}{n}}^{++}\cdots\tilde{\psi}_{-\frac{n-1}{2n}}^{++}|\tilde{\sigma}\rangle\ ,
\ee
where $|\tilde\sigma\rangle$ is the ground state of the $\mathbb{Z}_2$ twisted sector of the symmetric orbifold. We note that the fermions $\psi^{++}$ are integer-moded in the $\mathbb Z_2$ twisted sector. The total weight and R-charges then read:
\be\label{SymNS0Z2oddn_iv}
h=\bar{h}=\frac{1}{4n}+\frac{1}{8}\Big(n-\frac1n\Big)+\sum_{k=1}^{\frac{n-1}{2}} \left(\frac{k}{n} \right)=\frac{n}{4}\ ,\qquad\qquad\ell=\bar\ell=\frac{n-1}2+\frac12=\frac{n}2\ ,
\ee
which saturate the BPS bound. The $\mathbb Z_2$ twisted sector has multiplicity 2, so we obtain two BPS states from them, see eq. (\ref{S0Z2_bps}).

\subsubsection{even twisted sector of $\mathrm{Sym}^N(\mathcal S_0/\mathbb Z_2)$}\label{app_SymN_bps_even}
\textbf{${\bf{\mathbb Z_2}}$ untwisted sector:}

The ground-state energy of the even twist-$n$ sector of the symmetric orbifold is given by:
\be\label{SymNS0Z2evenn_i}
h=\bar h=\frac{(c_{\mathrm b}+c_{\mathrm f})n}{24}+\Big(-\frac{c_{\mathrm b}}{24n}+\frac{c_{\mathrm f}}{12n}\Big)=\frac{c_{\mathrm b}}{24}\Big(n-\frac1n\Big)+\frac{c_{\mathrm f}}{24}\Big(n+\frac2n\Big)\ ,
\ee
where $c_{\mathrm b}$ and $c_{\mathrm f}$ correspond to the boson and fermion central charges, see\cite[eq.s (A.5) and (A.9)]{Gaberdiel:2018rqv} and \cite[appendix D.2]{Eberhardt:2017pty}. In the even twisted sector of the symmetric orbifold, fermions have a different boundary condition than the bosons and their ground-state energies are consequently different \cite{Lunin:2001pw}. We have thus expressed the contributions from bosonic and fermionic fields separately in the above expression\footnote{We note that for $n=2$, the ground-state energies of a boson and an NS fermion are the same since $c_{\mathrm b}=1$ and $c_{\mathrm f}=\tfrac12$, hence the comment above equation (\ref{SymNS0Z2oddn_iii}). This, however, is not the case for $n>2$ in the even twisted sector of symmetric orbifold.\label{fn_neven}}.

For the $\mathcal S_0/\mathbb Z_2$ theory $c_{\mathrm b}=1$, $c_{\mathrm f}=\tfrac42=2$, and the ground state has
\be
h=\bar h=\frac 18\Big(n+\frac{1}{n}\Big)\ ,\qquad\qquad\ell=\bar\ell=\frac12\ .
\ee
Similar to the odd twisted sector analyses, we shall now apply fermionic excitation on the ground state to construct BPS states. However, the conformal dimension is too high to give BPS states in the $\mathbb{Z}_2$ untwisted sector, as described in \cite[section 3.3]{Eberhardt:2017pty}. Thus we obtain no BPS contributions in this case.

\vspace{10pt}\hspace{-25pt}
\textbf{${\bf{\mathbb Z_2}}$ twisted sector:}

We consider the state
\be\label{SymNS0Z2evenn_ii}
\psi_{-\frac{1}{2n}}^{++}\;\psi_{-\frac{3}{2n}}^{++}\cdots\psi_{-\frac{n-1}{2n}}^{++}\;
\tilde{\psi}_{-\frac{1}{2n}}^{++}\;\tilde{\psi}_{-\frac{3}{2n}}^{++}\cdots\tilde{\psi}_{-\frac{n-1}{2n}}^{++}|\tilde{\sigma}\rangle\ .
\ee 
It has conformal weight and $\mathfrak u(1)$-charge
\be\label{SymNS0Z2evenn_iii}
h=\bar h=\frac{1}{8}\left(n+\frac{1}{n}\right)-\frac{1}{8n}+\sum_{k=1}^{\frac{n}{2}} \frac{2k-1}{2n}=\frac{n}{4}\ ,\qquad\ell=\bar\ell=\frac{n-1}2+\frac12=\frac n2\ ,
\ee
which saturate the BPS bound. Here, the second term in the formula for the conformal weight is the ground-state energy of the twisted sector of the $\mathbb{Z}_2$-orbifold in the R sector divided by $n$, according to the general rule that a state of conformal weight $h$ in the seed theory contributes to conformal weight $\tfrac{h}{n}$ in the symmetric orbifold. Thus, we find two further BPS states in the even twisted sector of the symmetric orbifold. Taking all contributions together, we obtain the Hodge diamond \eqref{SymNS0Z2_bps}.

\subsection{Modified elliptic genus of the odd twisted sector of $\mathrm{Sym}^N(\mathcal S_0/\mathbb Z_2)$}\label{app_SymN_ellgen}
We derive the modified elliptic genus of the symmetric orbifold CFT in the odd twisted sector. Performing an S-modular transformation on the R-sector partition function of the seed theory with factors of the twist $n$ inserted, we obtain the single-particle $\widetilde{\rm NS}$ sector partition function of the twist-$n$ sector:
\begin{align}
\tilde Z_{\mathrm{NS}}(z,\tfrac{\tau}n;\bar z,\tfrac{\bar\tau}{n})&=\left|\frac{\vartheta_3(z|\tfrac\tau n)\vartheta_4(\tfrac\tau n)^{\frac{1}{2}}\vartheta_3(\tfrac\tau n)^{\frac{1}{2}}}{\vartheta_2(\tfrac\tau n)^{\frac{1}{2}}\eta(\tfrac\tau n)^{\frac{3}{2}}}\right|^2+\left|\frac{\vartheta_2(z|\tfrac\tau n)\vartheta_4(\tfrac\tau n)^{\frac{1}{2}}\vartheta_2(\tfrac\tau n)^{\frac{1}{2}}}{\vartheta_3(\tfrac\tau n)^{\frac{1}{2}}\eta(\tfrac\tau n)^{\frac{3}{2}}}\right|^2\label{Z_NSt_S0Z2_iv}\\
&=2\left|\frac{\vartheta_3(z|\tfrac\tau n)}{\vartheta_2(\tfrac\tau n)}\right|^2+2\left|\frac{\vartheta_2(z|\tfrac\tau n)}{\vartheta_3(\tfrac\tau n)}\right|^2\ ,\label{Z_NSt_S0Z2_v}
\end{align}
where we have not included the $\Theta$-dependent term since it contains $\theta_4(\tau)$ and vanishes after taking $\bar z=-\tfrac{\bar\tau}2$ in the modified elliptic genus.

Putting back the ground-state energy of the untwisted copies (\ref{SymNoddgs}),\footnote{This is the ground-state energy relative to the vacuum. We have not included the factor $q^{-\frac{c}{24}}$ in the partition function, since it diverges in the limit we are computing, see below eq.~\eqref{string elliptic genus definition}} we obtain
\bea\label{Z_NSt_sym_oddt}
&&{\tilde Z}_{\mathrm{NS}}(z,\tau;\bar z,\tau)=\sum_{n\in\mathbb Z_{>0,\mathrm{odd}}}
q^{\frac n8}\,\bar q^{\frac n8}{\tilde Z}_{\mathrm{NS}}(z,\tfrac\tau n;\bar z,\tfrac{\bar\tau}{n})\bigg|_{h-\bar h-\in\frac{\mathbb Z}{2}}\\
&&\qquad\qquad\qquad\;=\sum_{n\in\mathbb Z_{>0,\mathrm{odd}}} 2\,q^{\frac n8}\,\bar q^{\frac n8}\bigg(\left|\frac{\vartheta_3(z|\tfrac\tau n)}{\vartheta_2(\tfrac\tau n)}\right|^2+
\left|\frac{\vartheta_2(z|\tfrac\tau n)}{\vartheta_3(\tfrac\tau n)}\right|^2\bigg)\bigg|_{h-\bar h\in\frac{\mathbb Z}{2}}\nonumber
\eea
where we have included the untwisted sector contribution $n=1$ to write the general expression for all positive odd $n$. Invariance under the action of the $\mathbb Z_2$ imposes the constraint $h-\bar h\in{\mathbb Z}/{2}$. Using identities of the Jacobi theta functions which we list in appendix \ref{app_theta}, we find that the partition function reads
\bea\label{Z_NSt_sym_oddt_ii}
&&{\tilde Z}_{\mathrm{NS}}(z,\tau;\bar z,\tau)=\\
&&\sum_{n\in\mathbb Z_{>0,\mathrm{odd}}} 2\,q^{\frac n4}\,y^{\frac n2}\,\bar q^{\frac n4}\,\bar y^{\frac n2}\bigg(\left|\frac{\vartheta_2(z+\tfrac{\tau}2|\tfrac\tau n)}{\vartheta_2(\tfrac\tau n)}\right|^2
+\left|\frac{\vartheta_3(z+\tfrac{\tau}2|\tfrac\tau n)}{\vartheta_3(\tfrac\tau n)}\right|^2\bigg)\bigg|_{h-\bar h\in\frac{\mathbb Z}{2}}\ .\nonumber
\eea
We finally set $\bar h=\tfrac{\bar\ell}2$ to compute the modified elliptic genus:
\bea\label{ellgen_NSt_sym_oddt_iii}
\!\!\!\!\!\tilde{\mathcal{Z}}_{\mathrm{NS}}(z,\tau)=2\sum_{n\text{ odd}}
\bigg(q^{\frac n4}\,y^{\frac n2}\,\frac{\vartheta_2(z+\tfrac{\tau}2|\tfrac\tau n)}{\vartheta_2(\tfrac\tau n)}\Bigg|_{h\in \frac{\mathbb Z}{2}}+
q^{\frac n4}\,y^{\frac n2}\,\frac{\vartheta_3(z+\tfrac{\tau}2|\tfrac\tau n)}{\vartheta_3(\tfrac\tau n)}\Bigg|_{h\in \frac{\mathbb Z}{2}+\frac{1}{4}}\bigg)\ .
\eea

Following our approach in section \ref{CFT_ellgen_sym_odd}, we next perform a Fourier expansions of the modified elliptic genus and analyse the contributions from the $\mathbb Z_2$ untwisted and twisted sectors separately, see eq.s (\ref{ellgen_NSt_sym_oddt_iv}) and (\ref{ellgen_NSt_sym_oddt_v}). 

\vspace{10pt}
\hspace{-25pt}
\textbf{$\bf{\mathbb Z_2}$ untwisted sector}

Let us first consider the Fourier expansion of the $\mathbb Z_2$ untwisted sector in (\ref{ellgen_NSt_sym_oddt_iv}). For this, we define
\be 
2\frac{\vartheta_2(z|\tau)}{\vartheta_2(\tau)}\equiv\sum_{m \in \mathbb{Z},\, \ell \in \mathbb{Z}+\frac{1}{2}} c(m,\ell)q^m y^\ell\ .\label{c definition}
\ee
With this, we can expand the $\mathbb{Z}_2$ untwisted sector contribution as follows:
\bea
&\tilde{\mathcal{Z}}_{\mathrm{NS}}^{\mathrm{U_{\mathbb Z_2}}}(z,\tau)
&=\sum_{n\in\mathbb Z_{>0,\mathrm{odd}}}q^{\frac n4}\,y^{\frac n2}\sum_{m\in\frac{\mathbb Z}{n},\,\ell\in\mathbb Z+\frac{1}{2}} c(nm,\ell)\,q^{m+\frac \ell2}y^{\ell}\Big|_{h\in{\frac{\mathbb Z}{2}}}\label{ellgen_NSt_sym_oddt_vi}\\
&&=\sum_{n\in\mathbb Z_{>0,\mathrm{odd}}}q^{\frac n4}\,y^{\frac n2}\sum_{m\in\mathbb{Z},\,\ell\in\mathbb Z+\frac{1}{2}} c(nm,\ell)\,q^{m+\frac \ell2}y^{\ell}\ ,\label{ellgen_NSt_sym_oddt_via}\\
&&\!=\sum_{n\in\mathbb Z_{>0,\mathrm{odd}}}\sum_{m^\prime\in\frac{\mathbb Z}2,\,\ell^\prime\in\mathbb Z}
c\big(n(m^\prime-\tfrac{\ell^\prime}2),\ell^\prime-\tfrac n2\big)\,q^{\ell^\prime}y^{\ell^\prime}\ ,\label{ellgen_NSt_sym_oddt_vii}
\eea
where in the second line we have defined
\be\label{fouriers}
m^\prime\equiv m+\frac \ell2+\frac n4\ ,\qquad \ell^\prime\equiv \ell+\frac n2\ .
\ee
By restricting $m \in \mathbb{Z}$ in \eqref{ellgen_NSt_sym_oddt_via}, we have imposed the orbifold projection $h \in \frac{1}{2}\mathbb{Z}$.

The coefficients $c(m,\ell)$ are quasi-periodic in $z$ and satisfy the assumptions of Theorem 2.2 of \cite{eichler1985theory}:
\be\label{fouriers_ii}
c(m,\ell)\equiv c(2m-\ell^2)\ .
\ee
Eq. (\ref{ellgen_NSt_sym_oddt_vii}) then reads
\bea\label{ellgen_NSt_sym_oddt_viii}
&&\tilde{\mathcal{Z}}_{\mathrm{NS}}^{\mathrm{U_{\mathbb Z_2}}}(z,\tau)
=\sum_{n\in\mathbb Z_{>0,\mathrm{odd}}}\sum_{m^\prime\in\frac{\mathbb Z}2,\,\ell^\prime\in{\mathbb Z}} c\big(2nm^\prime-{\ell^\prime}^2-\tfrac{n^2}4\big)\,q^{m^\prime}y^{\ell^\prime}\nonumber\\
&&\qquad\qquad\;\;=\sum_{n\in\mathbb Z_{>0,\mathrm{odd}}}\sum_{m^\prime\in\frac{\mathbb Z}2,\,\ell^\prime\in{\mathbb Z}}
c\big(4{m^\prime}^2-{\ell^\prime}^2-(\tfrac{n}2-2m^\prime)^2\big)\,q^{m^\prime}y^{\ell^\prime}\nonumber\\
&&\qquad\qquad\;\;=\sum_{m^\prime\in\frac{\mathbb Z}2,\,\ell^\prime\in{\mathbb Z}}
\Bigg(\sum_{n\in\mathbb Z_{>0,\mathrm{odd}}}c\big(2{m^\prime}^2-\tfrac{{\ell^\prime}^2}2,\tfrac12(n-4m^\prime)\big)\Bigg)\,q^{m^\prime}y^{\ell^\prime}\ .
\eea
We next define $n^\prime\equiv n-4m^\prime$, which is an odd integer for $m^\prime\in\mathbb Z_{\ge0}/2$. The condition $n>0$ requires that $n^\prime>-4m^\prime$. We then first evaluate the sum over $n^\prime$ in eq. (\ref{ellgen_NSt_sym_oddt_viii}) and to do so, we use the fact that $c(s)=0$ for $s<-1$ and extend the range of the sum over all odd integers. This, however, may result in an over-counting of low-lying contributions which we then need to subtract subsequently, see \cite[section 5]{deBoer:1998us} and \cite[section 4.3]{Datta:2017ert}.We can evalute \eqref{ellgen_NSt_sym_oddt_viii} further using the definition \eqref{c definition}:
\be\label{ellgen_NSt_sym_oddt_x}
\sum_{n^\prime\in\mathbb Z_{\mathrm{odd}},\,s}c\big(s,\tfrac{n^\prime}2\big)q^s=
2 \frac{\vartheta_2(z|\tau)}{\vartheta_2(\tau)}\Bigg|_{z=0}=2\ ,
\ee
and hence
\be 
\sum_{n^\prime\in\mathbb Z_{\mathrm{odd}}}c\big(s,\tfrac{n^\prime}2\big)=2 \delta_{s,0}\ .
\ee
Thus, we find that
\be\label{ellgen_NSt_sym_oddt_xi}
\sum_{n^\prime\in\mathbb Z_{\mathrm{odd}}}c\big(2m^\prime-\tfrac{{\ell^\prime}^2}2,\tfrac{n^\prime}2\big)= 2\delta_{m^\prime,\pm\frac{\ell^\prime}2},\qquad \ell^\prime-2m^\prime\in\mathbb Z,\;|\ell^\prime|\ge1\ .
\ee
The only low-lying exception which we need to subtract from the above sum is for $\ell^\prime=0$. Inserting the result back in eq. (\ref{ellgen_NSt_sym_oddt_viii}), we find the expression \eqref{ellgen_NSt_oddSym_UZ2}.

\vspace{10pt}
\hspace{-25pt}
\textbf{$\bf{\mathbb Z_2}$ twisted sector}

The computation for the $\mathbb{Z}_2$ twisted sector is exactly analogous. The only differences arise in the summation ranges: now $m'\in \frac{1}{2}\mathbb{Z}+\frac{1}{2}$ and $\ell' \in \mathbb{Z}+\frac{1}{2}$. There are no low-lying exceptions in the twisted sector and we find the final result (\ref{ellgen_NSt_oddSym_TZ2}).

\section{Theta functions}\label{app_theta}
We follow the notation of \cite{Blumenhagen:2013fgp} and define the theta functions as
\be\label{app_theta_i}
\vartheta\bigg[
  \begin{array}{c}
  \alpha \\
  \beta
  \end{array}
\bigg](z|\tau)=\sum_{n\in\mathbb Z}e^{i\pi(n+\alpha)^2\tau+2\pi i(n+\alpha)(z+\beta)}\ .
\ee
The four Jacobi theta functions are then defined as
\be\label{app_theta_ii}
\vartheta_1\equiv\vartheta\bigg[
  \begin{array}{c}
  \frac12 \\
  \frac12
  \end{array}
\bigg]\ ,\qquad
\vartheta_2\equiv\vartheta\bigg[
  \begin{array}{c}
  \frac12 \\
  0
  \end{array}
\bigg]\ ,\qquad
\vartheta_3\equiv\vartheta\bigg[
  \begin{array}{c}
  0 \\
  0
  \end{array}
\bigg]\ ,\qquad
\vartheta_4\equiv\vartheta\bigg[
  \begin{array}{c}
  0 \\
  \frac12
  \end{array}
\bigg]\ .
\ee
We use the following identities of the theta functions:
\begin{align}
2\eta(\tau)^3&=\vartheta_2(\tau)\vartheta_3(\tau)\vartheta_4(\tau)\ ,\label{app_theta_iii_11}\\
\vartheta_2(z|\tfrac{\tau}n)&=q^{\frac n8}y^{\frac n2}\vartheta_3(z+\tfrac\tau2|\tfrac\tau2)\ ,\label{app_theta_iii_12}\\
\vartheta_3(z|\tfrac{\tau}n)&=q^{\frac n8}y^{\frac n2}\vartheta_2(z+\tfrac\tau2|\tfrac\tau2)\ ,\label{app_theta_iii_13}
\end{align}
where $q=e^{2\pi i\tau}$, $y=e^{2\pi iz}$, $\eta(\tau)$ is the Dedekind theta function and $n$ is an odd integer. The last two identities can be derived using eq.s (9.101b)-(9.101d) of \cite{Blumenhagen:2013fgp}.

\bibliographystyle{JHEP}
\bibliography{lN4N3}

\end{document}